\documentclass[sigconf]{acmart}

\usepackage{float}
\usepackage{graphicx}
\usepackage{colortbl}
\usepackage[normalem]{ulem}
\useunder{\uline}{\ul}{}
\usepackage{array}
\usepackage{tabularx}
\usepackage{booktabs}
\usepackage{lscape}
\usepackage{multirow, makecell}
\usepackage{rotating}

\AtBeginDocument{%
  }

\copyrightyear{2026}
\acmYear{2026}
\setcopyright{cc}
\setcctype{by}
\acmDOI{10.1145/3800645.3813054}

\acmConference[DIS ’26]{Designing Interactive Systems Conference}{June 13--17, 2026}{Singapore, Singapore}
\acmBooktitle{Designing Interactive Systems Conference (DIS ’26), June 13--17, 2026, Singapore, Singapore}
\acmISBN{979-8-4007-2563-0/2026/06}

\begin{document}

\title[AI Concept Envisioning Toolkit for Reflective Juxtaposition of Values and Harms]{Developing an AI Concept Envisioning Toolkit to Support Reflective Juxtaposition of Values and Harms}

\author{Pitch Sinlapanuntakul}
\orcid{0000-0003-3551-8531}
\affiliation{%
  \department{Human Centered Design \& Engineering}
  \institution{University of Washington}
  \city{Seattle}
  \state{WA}
  \country{USA}
}
\email{wspitch@uw.edu}

\author{Soyun Moon}
\authornotemark[1]
\orcid{0009-0003-3081-2392}
\affiliation{%
  \department{Human Centered Design \& Engineering}
  \institution{University of Washington}
  \city{Seattle}
  \state{WA}
  \country{USA}
}
\email{soyunm@uw.edu}

\author{Yuri Kawada}
\authornotemark[1]
\orcid{0009-0009-5311-1197}
\affiliation{%
  \department{Human Centered Design \& Engineering}
  \institution{University of Washington}
  \city{Seattle}
  \state{WA}
  \country{USA}
}
\email{ykawad@uw.edu}

\author{Yeha Chung}
\authornote{Authors contributed equally to this research.}
\orcid{0009-0008-1541-7569}
\affiliation{%
  \department{Human Centered Design \& Engineering}
  \institution{University of Washington}
  \city{Seattle}
  \state{WA}
  \country{USA}
}
\email{yehac@uw.edu}

\author{Mark Zachry}
\orcid{0000-0002-1067-7168}
\affiliation{%
  \department{Human Centered Design \& Engineering}
  \institution{University of Washington}
  \city{Seattle}
  \state{WA}
  \country{USA}
}
\email{zachry@uw.edu}

\renewcommand{\shortauthors}{Sinlapanuntakul et al.}

\begin{teaserfigure}
  \centering
  \includegraphics[width=\textwidth]{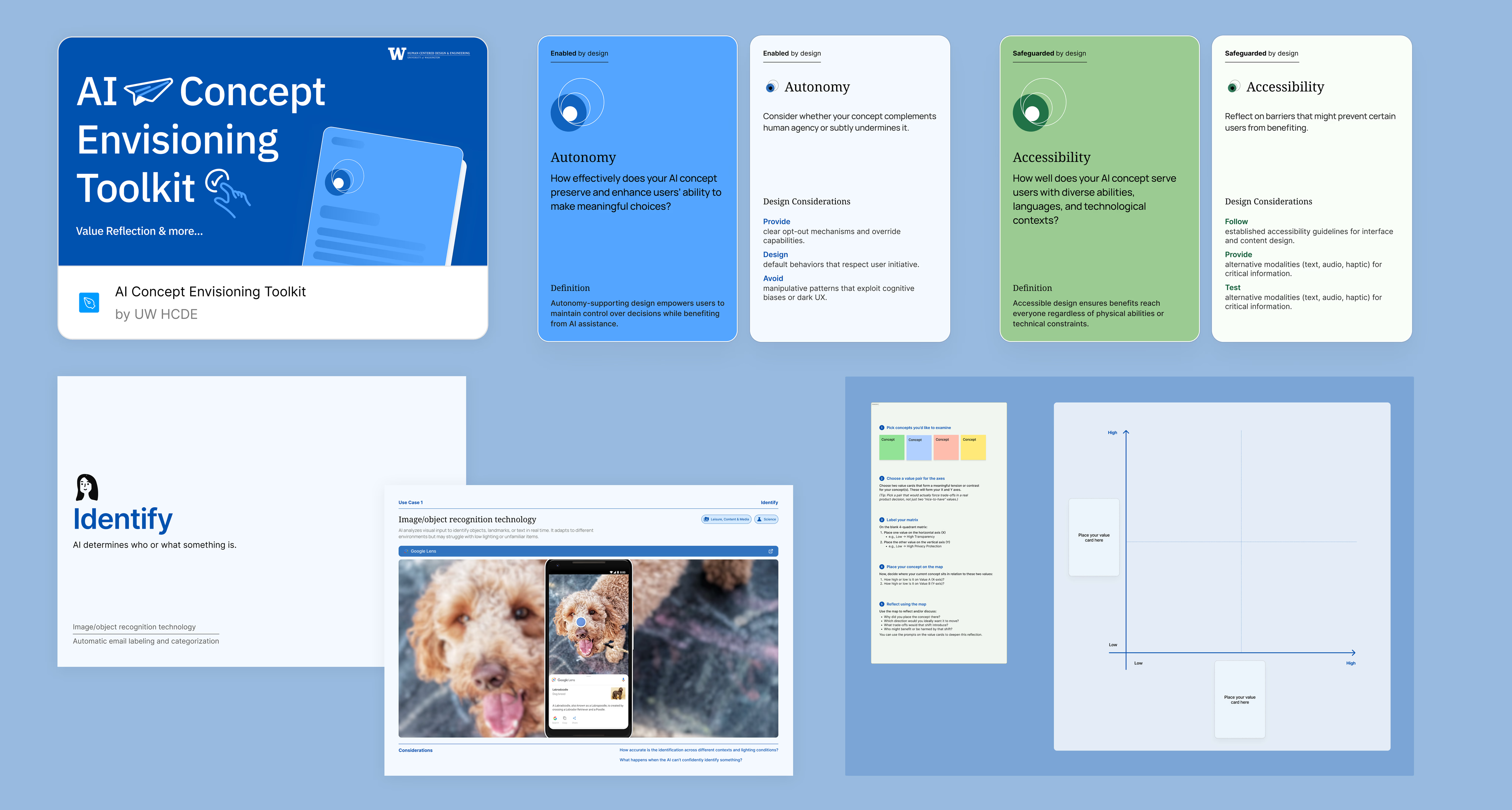}
  \caption{Final AI Concept Envisioning Toolkit with AI Capability Library, Value--Harm Cards, and Value--Tension Map.}
  \Description{Final AI Concept Envisioning Toolkit with AI Capability Library, Value--Harm Cards, and Value--Tension Map.}
  \label{fig:teaser}
\end{teaserfigure}

\begin{abstract}
Early-stage concept envisioning is a critical juncture in AI design, shaping how designers frame problems and the decisions that follow. Yet values and potential harms are often too abstract or addressed too late to meaningfully shape design. Using a Research-through-Design (RtD) approach, we developed the AI Concept Envisioning Toolkit, comprising an AI Capability Library, 24 Value--Harm Cards, and a Value--Tension Map, to support reasoning by juxtaposing values and harms within AI technical capabilities. Through a survey with 30 designers and in-depth interviews with 12 designers, we find that the toolkit is clear and perceived as valuable, and that it encourages value reflection, helps anticipate potential harms, and makes ethical considerations more transparent in early-stage design. We reflect on our design process and discuss design approaches for tools that promote reflection on values and potential harms, surface and navigate value tensions, and introduce productive friction throughout design workflows.
\end{abstract}




\maketitle

\section{Introduction}

The design of AI systems is fundamentally a process of translating human values into technical systems. Yet this translation routinely fails. AI-enabled products designed without explicit consideration of values and potential harms cause real damage. They disproportionately expose marginalized populations to risk, erode user autonomy, and reinforce systemic inequities in ways that often go unnoticed until deployment \cite{shelby2023, friedman2002, gabriel2020}. For instance, recommendation algorithms optimized for engagement distort how information reaches users, and educational AI designed to be helpful can quietly undermine student autonomy. These are not primarily technical mistakes but design failures rooted in decisions made early in concept envisioning \cite{saxena2025}. Such failures stem from choices about what the design should do, whose needs matter, and what risks are acceptable \cite{jung2025, namer2025harm96designers, shelby2023}. These choices often go unexamined until products or services are deployed and damage has occurred \cite{friedman2002, gabriel2020, whittlestone2019}. As AI becomes embedded in increasingly consequential domains, such as hiring and employment, healthcare diagnoses and treatment, criminal justice risk assessment, and education, the challenge of translating values into design becomes more urgent \cite{rahwan2019, shneiderman2020bridging, saxena2025, stray2024}. Yet design practice has not fundamentally shifted to address this challenge. Designers making foundational decisions about these technologies often do so without structured support for reasoning through value implications or anticipating harms. Instead, they rely on intuition, organizational priorities, or technical feasibility \cite{buchanan1992, yang2018UXML, holstein2019}. This temporal misalignment, where values matter most but are supported least, represents a critical gap in current design practice and tools \cite{morley2020, wong2023}.

Early-stage concept envisioning is the critical juncture where designers have the greatest leverage to shape value-aligned outcomes, but they lack integrated frameworks and tools calibrated to this specific phase. Existing AI ethics and design resources provide valuable guidance but share critical limitations \cite{googlepair, microsoftguidelines, morley2020}. Many assume prior AI literacy and position ethical reasoning primarily as late-stage evaluation rather than as generative ideation \cite{sadek2024designing, wong2023}. Prior research shows that designers struggle to translate abstract values, like autonomy or fairness, into concrete AI behaviors and feature-level decisions \cite{friedman2019, nishal2025, sinlapanuntakul2025designjam, sinlapanuntakul2026envision}. They also rarely pause to uncover value tensions or reason about long-term and cumulative harms, particularly during fast-paced ideation \cite{whittlestone2019}. Compounding this, designers typically ideate within tools such as Figma, FigJam, or Miro, while ethics-related resources exist elsewhere, creating context-switching costs that exceed perceived benefit at moments of decision \cite{yildirim2023designresources, wong2023}. This gap between where ethical reasoning is needed and where it is available represents a persistent structural challenge for any tool operating in this space, not a failure unique to prior work. Together, these conditions lead to late-stage consideration of ethics, when technical commitments have solidified and redesign becomes costly \cite{sadek2024designing, morley2020}.

Our research asks: (i) \textit{How might we create a toolkit that helps designers deliberately reflect on values and potential harms during early-stage AI concept development?} and (ii) \textit{How does such a toolkit support harm-aware value reflection while designers are envisioning AI concepts?} To address these questions, we adopted a research-through-design (RtD) approach \cite{zimmerman2014RtD}, conducting a two-phase mixed-methods study with design professionals. In Phase 1, we developed a standalone web-based toolkit and evaluated it with 30 designers. While participants valued the content, they reported adoption friction due to its separation from everyday workflows and limited support for collaborative manipulation. In Phase 2, we redesigned the toolkit into FigJam\footnote{https://www.figma.com/figjam/} for the Figma Community and conducted in-depth studies with 12 designers using a design activity, think-aloud protocols, and interviews. The redesigned AI Concept Envisioning Toolkit demonstrates that value-based reflection can be meaningfully integrated into early-stage AI design when tools are embedded within existing workflows, structured around value--harm juxtaposition, and designed to introduce productive friction rather than seamless automation \cite{schon1983, giaccardi2015}.

Our analysis reveals three key insights. First, spatially juxtaposing conflicting values prompts designers to confront tensions they might otherwise rationalize away, supporting reflective conversation with design materials \cite{schon1992, goldschmidt1991}. Second, early encounters with potential harms reshape the design problem itself rather than merely adding constraints, aligning with prior work on framing and problem setting in design \cite{dorst2015, nelson2014}. Third, toolkit coherence matters more than any individual component, as designers move fluidly between capability exploration, value--harm cards, and value--tension mapping. Beyond the toolkit artifact, we contribute conceptually by showing that tool effectiveness depends not only on content but on how tools are distributed, integrated, and materialized \cite{wong2023, zimmerman2014RtD}; methodologically by demonstrating how RtD generates insights about toolkit development in practice; and practically by providing a freely accessible Figma Community resource\footnote{https://bit.ly/ACEtoolkit} distributed to the broader HCI and design practice community. The temporal and spatial alignment between where ethical reflection is needed and where it is available shapes designer engagement, while the material affordances of tools determine what forms of reasoning become possible. These insights extend beyond AI design to any domain where practitioners require structured, situated support for reasoning about values in everyday practice.


\section{Background and Related Work}

\subsection{Challenges of Designing AI Concepts}

The integration of AI into products and services raises ethical questions that are inseparable from the technical design process. Decisions about data sources, algorithmic objectives, interaction patterns, and feedback loops inherently encode assumptions about what matters and whose needs take priority, making every technical choice a value-laden one. Scholarship in value-sensitive design \cite{friedman2019}, design theory \cite{doordan2003, dorst2015}, and the politics of artifacts \cite{scheuerman2021, rifat2024} collectively underscores that design decisions are not neutral, as the framing of problems shapes which solutions are imaginable, and artifacts themselves encode particular worldviews and priorities. In AI systems, these dynamics are amplified because emergent system behavior depends on training data, model architecture, and deployment context \cite{dove2017, holmquist2017}. Despite this, empirical studies on early-stage AI design consistently show that ethical and value considerations are rarely made explicit during concept ideation \cite{li2024, yildirim2024}. Designers often rely on intuition, organizational pressures, or technical feasibility to guide decisions, \cite{sinlapanuntakul2025SLR} and when values are considered, it is typically as a downstream validation exercise rather than as a generative influence on design concepts. Early engagement of diverse, interdisciplinary design perspectives in AI projects is important \cite{holstein2019, nahar2022, deng2023, yildirim2022}, yet this is precisely the phase with the fewest tools to support designers when they are most receptive to questioning assumptions and exploring alternatives \cite{dorst2015, buchanan1992}.

Early-stage concept envisioning is precisely the point at which designers hold the most leverage over downstream outcomes, but it is also the stage with the least structured support \cite{jung2025, saxena2025}. Once design concepts solidify, organizational momentum and technical commitments make significant changes costly or unlikely, while the lack of scaffolding during ideation means that value and harm considerations are often deferred until it is too late to influence system behavior meaningfully \cite{namer2025harm96designers}. Existing frameworks tend to assume teams with established AI literacy and narrowly scoped problems, leaving the open, exploratory, and inherently ambiguous space of early ideation under-supported \cite{liao2023, yang2020}. The resulting mismatch between when ethical reasoning matters most and when it is actually facilitated has been widely documented and constitutes a persistent gap in both research and practice.


\subsection{Toolkits in Design Practice}

Design-led toolkits have long been used to support design practitioners in activities like ideation and collaborative reasoning. Early contributions, such as “design by playing” \cite{ehn1988} and experience prototyping \cite{buchenau2000}, illustrate how mock-ups and tangible artifacts allow designers to explore scenarios in ways that are discussable, testable, and manipulable. Toolkits have since proliferated across multiple domains, from service design \cite{stickdorn2012} to speculative design \cite{dunne2013} and inclusive design \cite{abascal2005, gilbert2025}. Across these studies, a consistent insight emerges that tool adoption and effectiveness are contingent on alignment with existing workflows \cite{subramonyam2022, muralikumar2025}. In particular, designers engage more readily with resources embedded in familiar contexts. Context-switching imposes cognitive and practical friction that reduces sustained engagement \cite{knearem2023}. Schön’s concept of reflection-in-action \cite{schon1983} further emphasizes that learning and insight occur in the moment of doing; tools that are external or isolated from practice risk being ignored or superficially applied.

Several existing resources address designing AI with ethics \cite{jung2025, wilson2024, sadek2024guidelines, saxena2025, park2025, googlepair, ibmguidelines, microsoftguidelines, weisz2024, yildirim2023designresources}, but all share limitations in supporting early-stage value reasoning. For example, Google’s People + AI Guidebook \cite{googlepair} provides UX patterns for AI systems but assumes prior AI literacy and is oriented toward implementation rather than concept exploration \cite{yildirim2023pairguidebook}. Microsoft’s Guidelines for Human-AI Interaction focuses on human-centered evaluation post-design \cite{microsoftguidelines}, while the AI Brainstorming Kit, created by HCI researchers at Carnegie Mellon University, provides a taxonomy of AI capabilities without explicit scaffolding for reasoning about values or potential harms \cite{yildirim2023designresources}. Recent work on AI design resources and toolkits \cite{wong2023, feng2023} reveals three notable shortcomings in existing approaches. First, they are largely external to workflow, existing as PDFs, standalone websites, or workshop materials that require context-switching. Second, they emphasize late-stage evaluation rather than generative reflection during early ideation, positioning ethical reasoning as a compliance check rather than a creative force \cite{liao2020, morley2020}. Third, their prioritization of clarity and linear workflows may inadvertently suppress engagement with ambiguity. Early-stage design is inherently uncertain and multi-dimensional \cite{nelson2014, cross2021}, yet many tools simplify these conditions, constraining designers’ ability to explore ethical trade-offs, anticipate harms, or co-construct meaning through iterative, reflective practice. Collectively, these gaps highlight the need for resources that integrate into daily workflows, support situated reflection, and leverage ambiguity as a productive force rather than something to be minimized.


\subsection{Reflection, Ambiguity, and Research-through-Design}

Reflection is a central mechanism through which practitioners develop understanding and make intentional choices. Schön \cite{schon1983} describes reflection-in-action as an iterative process in which thinking and making co-occur, with tangible artifacts providing feedback that shapes understanding. Subsequent work demonstrates that structured prompts, visual scaffolds, and collaborative discussion amplify both individual and collective insight \cite{wong2023, shi2023}. Importantly, reflection in early-stage design is exploratory rather than problem-solving. It is a mechanism for sense-making, negotiating ambiguity, and surfacing assumptions rather than identifying a single \textit{correct} solution. Design problems are often \textit{wicked}, characterized by incomplete information, conflicting objectives, and evolving requirements \cite{buchanan1992, cross2021}. In the design of AI-enabled solutions, this complexity is compounded because values, potential harms, and stakeholder needs are seldom considered in advance \cite{abedin2022}. Reasoning about them requires iterative negotiation, visualization, and discussion across cycles of making and reflection. Tools that reduce these challenges to checklists, numeric scales, or rigid workflows risk constraining the very reasoning they are meant to support \cite{morley2020, shen2024bidirectional}.

Effective design tools must do more than convey information; they must actively shape reasoning processes. Research on design support shows how well-designed constraints can disrupt habitual thinking, prompting designers to interrogate assumptions, weigh competing values, and anticipate potential harms \cite{giaccardi2015, moussette2010}. This work suggests that reflection is not simply a cognitive activity but emerges through interaction with structured material and spatial arrangements, which create opportunities for reconsideration that would not occur in unstructured ideation. RtD further reinforces this perspective by showing that knowledge about design practice emerges not only from final artifacts, but also from designers’ engagement with the tools and scaffolds that structure their work \cite{zimmerman2014RtD}. Studies on tool adoption demonstrate that effectiveness commonly depends on integration into practice \cite{subramonyam2022, kross2021, muralikumar2025}. Toolkits that are manipulable, situated in workflow, and capable of supporting iterative, reflective action are more likely to produce meaningful insight than prescriptive or static guidance. Taken together, while existing toolkits provide capability taxonomies, pattern libraries, or ethical checklists \cite{jansen2023, knearem2023}, they rarely structure conditions that make ambiguity and ethical tension generative, nor do they embed reflection in ways that are directly actionable during early-stage concept development.


\section{Toolkit Design as Inquiry}

We developed the toolkit through iterative cycles of sketching, storyboarding, piloting, and co-reflection, treating the design process itself as a method of inquiry \cite{zimmerman2014RtD}. Each iteration surfaced key questions about how to support designers in imagining AI possibilities while reflecting on human values and potential harms. Following Greenberg and Buxton \cite{greenberg2008}, our early design evaluations employed practices akin to traditional design critiques, where designers presented the design (i.e., toolkit) to the team, articulated their design rationale, and received constructive feedback. This interactive process allowed both presenters and critics to surface assumptions and explore alternatives in real time, embodying the reflective conversation model central to design practice \cite{schon1992}.

Grounded in this reflective RtD approach \cite{zimmerman2014RtD}, our toolkit builds on conceptual foundations from value-focused design frameworks \cite{friedman2019, friedman2013, sadek2024designing, shneiderman2020bridging, umbrello2021, zhu2018} and the tradition of conceptual investigation exemplified by the Envisioning Cards \cite{envisioningcards, friedman2019}. The Envisioning Cards organize reflection around four criteria (i.e., Stakeholder, Time, Value, and Pervasiveness) that prompt designers to consider who is affected, across what timeframes, according to which values, and at what scales of deployment \cite{envisioningcards, yurrita2022}. Our toolkit shares this commitment to early-stage value reflection but makes a distinct move. Instead of asking designers to \textit{identify} values and stakeholders, our toolkit asks designers to \textit{collide} values with harms.

Placing a value card adjacent to a harm card, within the context of a specific AI capability, generates productive friction that foregrounds trade-offs designers would otherwise rationalize away. This juxtaposition is further structured by our value--harm categorization. Rather than presenting human values as a flat repertoire, we organize them as either Enabled-by-Design or Safeguarded-by-Design, introducing directionality that requires designers to reason about values as a relational system rather than a checklist \cite{umbrello2021, shelby2023, liao2020, wong2023}. The Value--Tension Mapping then externalizes these collisions spatially, turning abstract trade-offs into a manipulable design artifact that supports negotiation \cite{schon1983, giaccardi2015}. Other card- or game-based methods, such as Judgment Call \cite{ballard2019}, similarly engage designers in benefit-harm deliberation; our contribution lies not in introducing ethical conversation to design but in matchmaking \cite{bly1999matchmaking}, bringing value--harm tensions into contact with AI capabilities at the moment of ideation, within the design environments practitioners already use \cite{morley2020}, such as FigJam.


\subsection{Toolkit Components (version 1)}

The first iteration of the toolkit consists of two main components, the AI Capability Library and the Value--Harm Cards, which can be used flexibly throughout concept design. This dual scaffolding supports reasoning about both what an AI can do and what it should do. By turning abstract reflection into tangible, iterative practice, the toolkit helps designers anticipate harms, surface value tensions, and engage critically and interpretively during ideation \cite{abedin2022}. Its nonlinear structure lets designers begin with AI capabilities, values, or specific scenarios, cycling between them as needed. This flexibility mirrors the situated and iterative nature of value reflection, especially in concept envisioning \cite{schon1983, buxton2010}. A third component, the Value--Tension Map, was introduced as a direct result of Phase 1 findings and is described in \hyperref[sec:valuetensionmap]{Section~\ref*{sec:valuetensionmap}} alongside the design rationale that motivated it.

\begin{figure}[H]
    \centering
    \includegraphics[width=\linewidth]{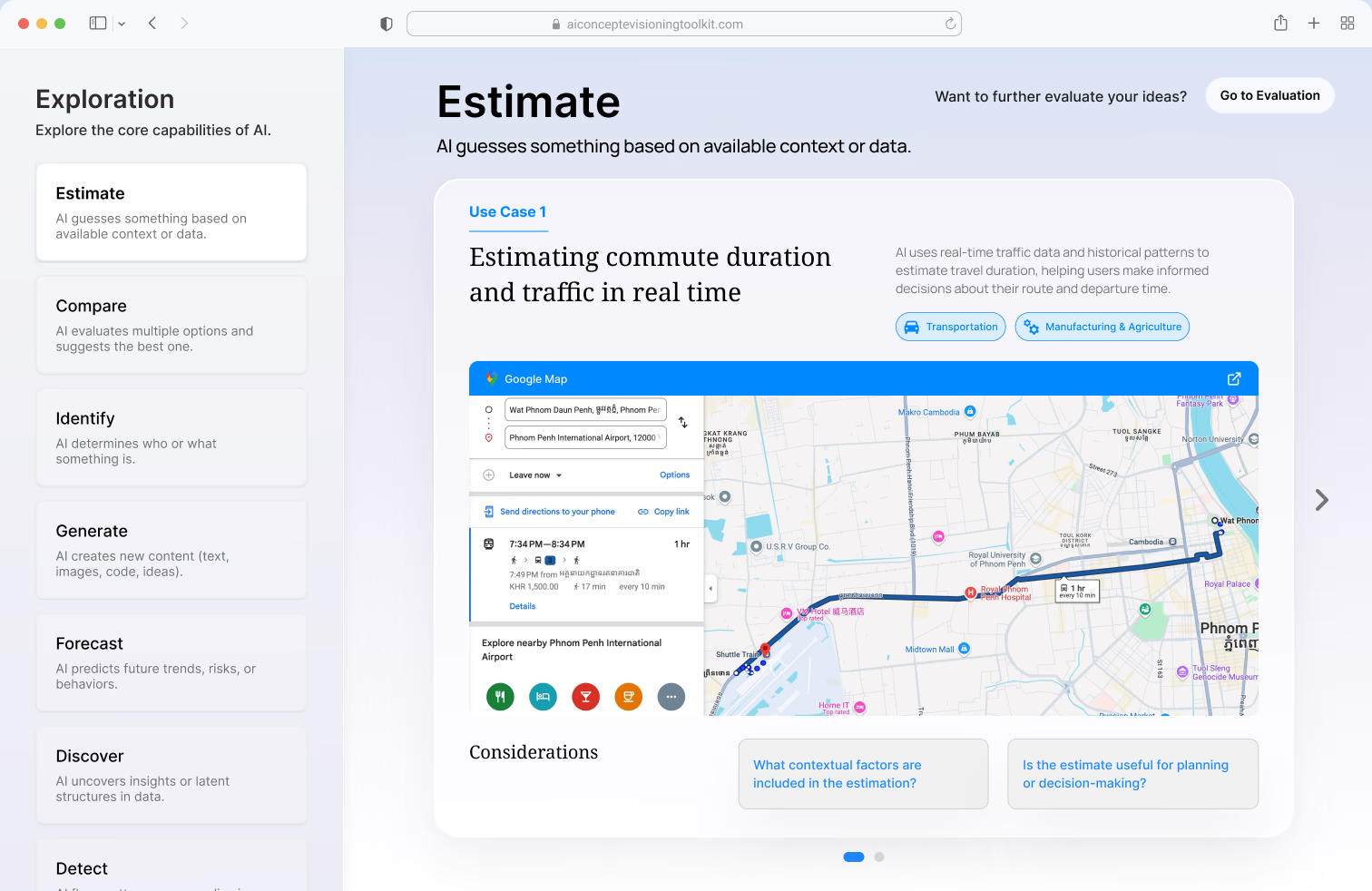}
    \caption{Example of AI Capability Library (version 1).}
    \Description{Example of AI Capability Library (version 1) on a web-based interface displaying 8 capabilities on the left and the “Estimate” capability selected, showing its use case with visuals, descriptions, and suggested design considerations.}
    \label{fig:toolkit-v1-library}
\end{figure}

\paragraph{AI Capability Library} The library presents eight foundational AI capabilities: act, compare, detect, discover, estimate, forecast, generate, and identify, each paired with two concrete, domain-specific use cases. As illustrated in \hyperref[fig:toolkit-v1-library]{Figure~\ref*{fig:toolkit-v1-library}}, every example includes a brief description of the capability, real-world applications, and design considerations highlighting opportunities and constraints. For instance, the \textit{Discover} capability illustrates content recommendation scenarios in productivity and health contexts, explaining how AI analyzes user behavior and surfaces relevant or unexpected content. 

The library was adapted from the AI Brainstorming Kit \cite{yildirim2023designresources} and optimized for designers. It is visually oriented, emphasizes quick access to information, and frames functionality through familiar real-world examples, such as Netflix-style recommendations, rather than conceptual descriptions. These design choices reduce the blank page problem \cite{crilly2019}, helping designers reason simultaneously about technical possibilities, interaction patterns, and societal implications within a given design context \cite{yang2018UXML, kayacik2019}.

\paragraph{Value--Harm Cards} Grounded in value-focused and AI ethics literature \cite{friedman2019, umbrello2021, sadek2024guidelines}, the 24 cards are organized into two harm--aware categories: 14 \textit{Enabled-by-Design} cards and 10 \textit{Safeguarded-by-Design} cards. The initial value list was identified through a synthesis of VSD’s standard value repertoire \cite{friedman2019}, AI ethics literature \cite{sadek2024guidelines, sadek2024designing, sinlapanuntakul2026envision}, and harm taxonomy frameworks \cite{shelby2023, umbrello2021}. We iteratively refined the list by consolidating overlapping constructs and removing values that were either too abstract or too narrowly technical to support actionable design reasoning. The final selection of values was further shaped through multiple discussions with researchers, design practitioners, and a human-centered AI design expert. The construction of individual cards, including how each value was defined and framed, was then refined using research through co-design \cite{busciantella2024RtC} and playtesting \cite{fullerton2004playtesting} sessions. Early iterations that framed values directly from ethical frameworks proved too abstract, surfacing the need for contextualized prompts and ripple-effect questions. Inspired by Umbrello and Van de Poel \cite{umbrello2021}’s distinction between \textit{Promoted by Design} and \textit{Respected by Design}, we categorized our final set of values as either \textit{Enabled-by-Design} (i.e., supporting positive human experiences) or \textit{Safeguarded-by-Design} (i.e., protecting against harmful experiences) based on whether their primary design orientation is generative or protective.

\begin{figure}[H]
    \centering
    \includegraphics[width=\linewidth]{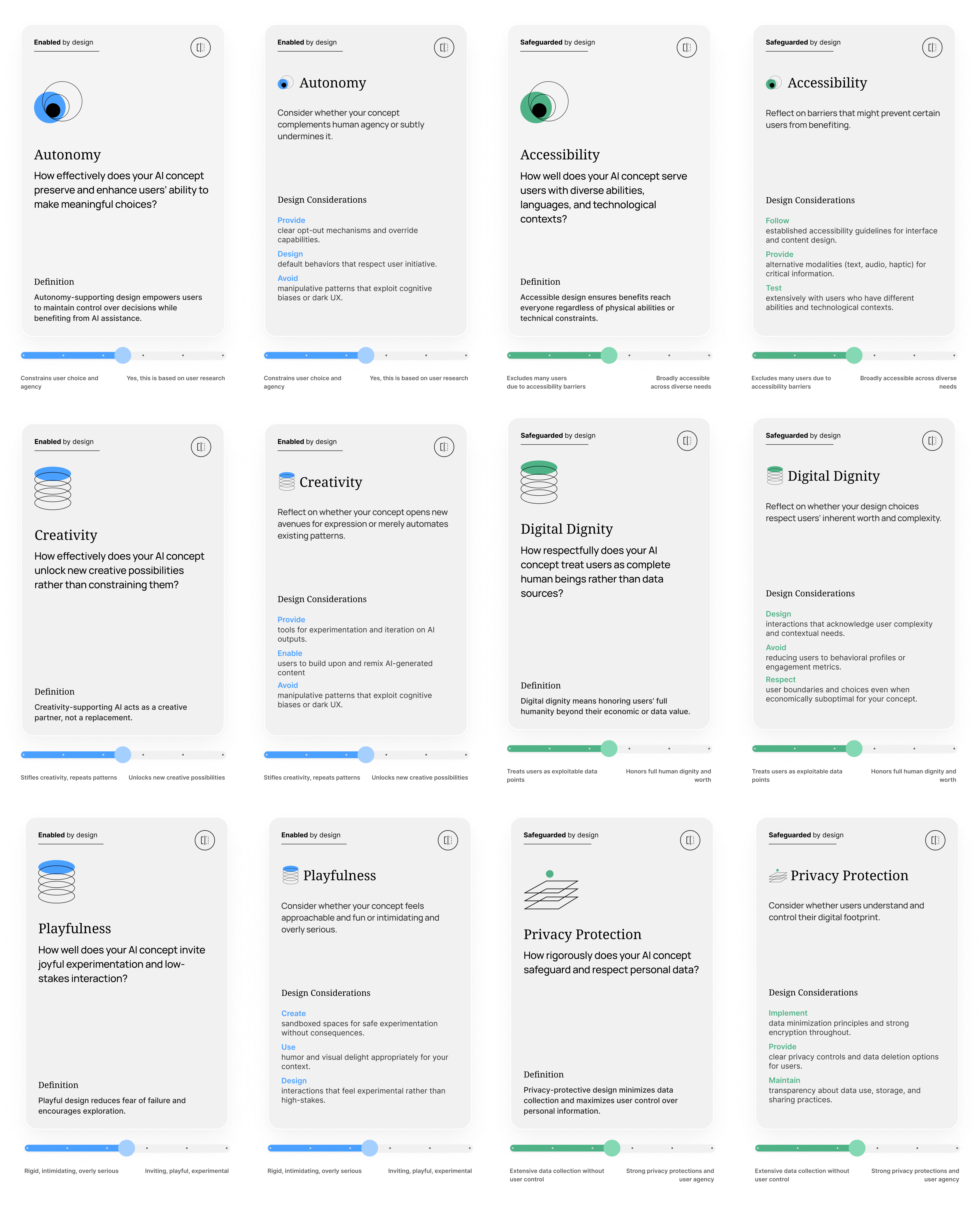}
    \caption{Examples of Value--Harm Cards (version 1).}
    \Description{Examples of Value--Harm Cards (version 1), showing the front and back of 6 of the cards with each card including its value principle, reflective prompt, ripple-effect questions, definitions, design considerations, and quantifiable sliders for concept evaluation.}
    \label{fig:toolkit-v1-cards}
\end{figure}

As shown in \hyperref[fig:toolkit-v1-cards]{Figure~\ref*{fig:toolkit-v1-cards}}, each card contains a concise value definition, reflective prompts linking the value to design choices, real-world examples, and an interactive seven-point rating slider (later removed in version 2; see \hyperref[sec:designiterations]{Section~\ref*{sec:designiterations}}). Designers can pair and group cards to explore tensions, evaluate evolving concepts, and navigate between generative and protective reasoning. This design encourages iterative reflection across personal, interpersonal, and societal levels, enabling designers to actively negotiate values rather than passively follow rules \cite{nishal2025, nakata2025}. The card-based interface supports rapid manipulation, aligning with natural design practices while maintaining interpretive flexibility \cite{giaccardi2015, moussette2010, hsieh2023cards}.


\subsection{Reflection}

Central to these conceptual directions is our position of reflection as a reasoning-centered activity rather than a checklist \cite{wong2023, liao2020} and our intentional exclusion of AI from the reflective process itself \cite{sinlapanuntakul2026envision}. While AI tools can shape designers’ thinking \cite{fu2024, palani2024}, overreliance risks outsourcing the critical and creative judgments that matter most, such as which problems deserve attention, which values should guide design, and what harms may emerge \cite{popa2021, gabriel2020}. By providing structured, AI-independent scaffolds, our toolkit supports designers in developing their own reasoning patterns while enabling informed exploration of AI possibilities \cite{yang2020, liao2023}. This keeps reflection grounded in designers’ judgment rather than in the implicit assumptions embedded in AI-generated outputs, grounding our work in principles of human-centered AI design \cite{shneiderman2020hcai, xu2019}.

In developing the toolkit, we came to see AI less as a technical black box to be explained and more as a material to be shaped and interpreted \cite{dove2020, holmquist2017}. This shift prompted us to move our focus from \textit{how AI works} to \textit{what AI does to people}, a distinction that became crucial when designing for designers who were less concerned with technical mechanics but deeply attuned to interaction and meaning. Our early prototypes leaned heavily on definitions and abstract principles, but we found that these tended to constrain thinking. Through design critiques, we noticed that introducing contextual prompts and leaving room for ambiguity allowed for deeper engagement on what a value might mean in their specific design scenario \cite{tian2024}. Over time, the toolkit evolved from what felt like educational material to what we now see as a set of critical provocations. Reflection became richer when designers could question, negotiate, and reinterpret values, rather than simply accept them as fixed guidelines \cite{benjamin2021, nishal2025}. This process revealed a persistent tension between how much guidance to provide without limiting interpretive freedom. Ultimately, we realized that supporting reflection is about creating space, with subtle nudges, for designers to explore, experiment, and critically engage with the consequences of their choices.


\section{Phase 1: Survey}

Phase 1 assessed the extent to which the toolkit was conceptually clear and valuable to designers, using a survey of 30 design professionals with current or past experience designing AI products or services.


\subsection{Method}

This study was approved by our University’s Institutional Review Board (IRB). In Phase 1, we conducted a mixed-methods survey evaluation combining structured quantitative ratings and open-ended written responses to assess toolkit usefulness, usability, and adoption barriers among design professionals (see the Survey Protocol in \hyperref[appendix:surveyprotocol]{Appendix~\ref*{appendix:surveyprotocol}}). This pairing enabled us to capture both the toolkit’s conceptual value and the real-world barriers practitioners faced when considering adoption.

\begin{table*}[t]
    \caption{Survey participants’ demographic information (S01–S30).}
    \Description{Table displaying self-reported demographic data of 30 survey study participants including gender, age, years of design experience, primary role, and organizational size.}
    \label{tab:participants-survey}
    \centering
    \small
    \begin{tabularx}{0.7\textwidth}{
        >{\raggedright\arraybackslash}p{1cm}  
        >{\raggedright\arraybackslash}p{1.5cm} 
        >{\raggedright\arraybackslash}p{1cm}  
        >{\raggedright\arraybackslash}p{1.5cm} 
        >{\raggedright\arraybackslash}p{3.7cm} 
        >{\raggedright\arraybackslash}X  
    }
    \toprule
    \textbf{ID} & 
    \textbf{Gender} & 
    \textbf{Age} & 
    \textbf{Exp. (yrs)} & 
    \textbf{Primary Role} & 
    \textbf{Org Size} \\
    \midrule
    S01 & Female & 38 & 14 & UX / Product Designer & >10k \\
    S02 & Male & 29 & 6 & UX / Product Designer & 100–10k \\
    S03 & Female & 33 & 9 & UX / Product Designer & 100–10k \\
    S04 & Female & 37 & 15 & Design Researcher & < 100 \\
    S05 & Male & 24 & 3 & Interaction Designer & 100–10k \\
    S06 & Male & 51 & 25 & UX / Product Designer & >10k \\
    S07 & Female & 31 & 7 & UX / Product Designer & < 100 \\
    S08 & Female & 34 & 12 & Design Researcher & 100–10k \\
    S09 & Male & 27 & 4 & Interaction Designer & 100–10k \\
    S10 & Female & 35 & 12 & UX / Product Designer & 100–10k \\
    S11 & Female & 26 & 3 & UX / Product Designer & < 100 \\
    S12 & Male & 39 & 16 & UX / Product Designer & >10k \\
    S13 & Female & 30 & 5 & UX / Product Designer & 100–10k \\
    S14 & Male & 28 & 5 & UX / Product Designer & < 100 \\
    S15 & Female & 48 & 22 & UX / Product Designer & >10k \\
    S16 & Female & 40 & 16 & Interaction Designer & 100–10k \\
    S17 & Male & 32 & 8 & UX / Product Designer & >10k \\
    S18 & Female & 23 & 1 & Interaction Designer & 100–10k \\
    S19 & Female & 35 & 10 & UX / Product Designer & >10k \\
    S20 & Male & 29 & 6 & UX / Product Designer & < 100 \\
    S21 & Female & 26 & 3 & Interaction Designer & 100–10k \\
    S22 & Male & 44 & 22 & Design Researcher & >10k \\
    S23 & Female & 27 & 4 & UX / Product Designer & < 100 \\
    S24 & Male & 34 & 9 & Design Researcher & 100–10k \\
    S25 & Female & 43 & 18 & UX / Product Designer & >10k \\
    S26 & Male & 36 & 10 & Interaction Designer & >10k \\
    S27 & Male & 29 & 6 & UX / Product Designer & < 100 \\
    S28 & Male & 31 & 7 & Design Researcher & 100–10k \\
    S29 & Female & 25 & 3 & UX / Product Designer & < 100 \\
    S30 & Male & 33 & 11 & Design Researcher & 100–10k \\
    \bottomrule
    \end{tabularx}
\end{table*}

\subsubsection{Participants}

Thirty design and product professionals participated in the survey study (see \hyperref[tab:participants-survey]{Table~\ref*{tab:participants-survey}}). Eligibility criteria required participants to hold a UX/product design, interaction design, or design research role and to have worked on an AI-enabled product as a designer within the previous two years. Participants were recruited through diverse professional design networks, including X (formerly Twitter), LinkedIn, and HCI- and design-specific communities on Slack and Discord. A brief pre-screening questionnaire confirmed role, AI project experience, and organizational context before participants were invited. Participants included UX/Product Designers (\textit{n} = 18), Interaction Designers (\textit{n} = 6), and Design Researchers (\textit{n} = 6). Professional experience ranged from 3 to 25 years (\textit{M} = 9.73, \textit{SD} = 6.35), spanning junior-level to senior design leadership. Organizationally, participants represented small agencies (\textit{n} = 8, <100 employees), mid-size firms (\textit{n} = 13, 100--10k employees), and large enterprises (\textit{n} = 9, >10k employees) across diverse sectors (i.e., fintech, healthcare, media, education, and consumer technology).

\subsubsection{Procedure}

Participants watched two walkthrough videos, one for each main toolkit component: the AI Capability Library and the Value--Harm Cards. After each walkthrough, they rated each component across three evaluation dimensions (\textit{perceived clarity}, \textit{usefulness}, and \textit{comprehensiveness}), using 7-point Likert scales with anchored endpoints (e.g., 1 = Not at all useful; 7 = Very useful). They also rated \textit{overall usefulness} and \textit{likelihood to adopt} for the toolkit as a whole on the same scale. All five measures were operationalized as single-item scales to minimize participant burden and maintain ecological validity for an evaluative survey. Open-ended responses captured how each component could support participants’ work, optimal usage contexts within their processes, perceived advantages over current approaches or practices, barriers to adoption, and suggestions for improvement. These questions allowed participants to contextualize ratings, reveal adoption barriers not evident in scales, and clarify alignment between toolkit features and their actual workflows.

\subsubsection{Analysis}
We calculated descriptive statistics for all rating dimensions. Paired-samples t-tests compared \textit{AI Capability Library} and \textit{Value--Harm Cards} ratings across clarity, usefulness, and comprehensiveness; effect sizes (Cohen’s d) were reported to indicate magnitude. Paired-samples t-tests were used because each participant evaluated both components, creating a within-subjects structure in which the same individual provided ratings across all three dimensions. These comparisons are descriptive rather than evaluative in that they characterize the relative perceived strengths of the two components to motivate and justify specific design iterations (in \hyperref[sec:designiterations]{Section~\ref*{sec:designiterations}}), rather than to assess toolkit effectiveness against a baseline condition. Component order was not randomized to preserve a typical experience of navigating the toolkit (version 1). No causal claims about toolkit impact were drawn from Phase 1 data alone. Pearson correlations examined associations between component ratings and overall toolkit impressions. 

Prior to analysis, all 30 responses were inspected for completeness and screened for invariant response patterns; no responses were excluded on either basis. The uniformly high ratings observed are interpreted as consistent with the purposively composed sample of experienced, AI-engaged design professionals, acknowledged in \hyperref[sec:limitations]{Section~\ref*{sec:limitations}}. Two researchers analyzed open-ended responses using affinity diagramming \cite{hanington2019, karen2017}, iteratively generating codes reflecting toolkit strengths, barriers to adoption, and design recommendations, which were then synthesized into high-level themes. Discrepancies throughout the process were discussed until the two researchers reached consensus. Themes were cross-checked against quantitative findings to identify convergence and tensions between ratings and open-ended responses. Representative quotes were selected to illustrate key themes and ground the findings.


\subsection{Findings}

\begin{figure}[H]
    \centering  
    \includegraphics[width=\linewidth]{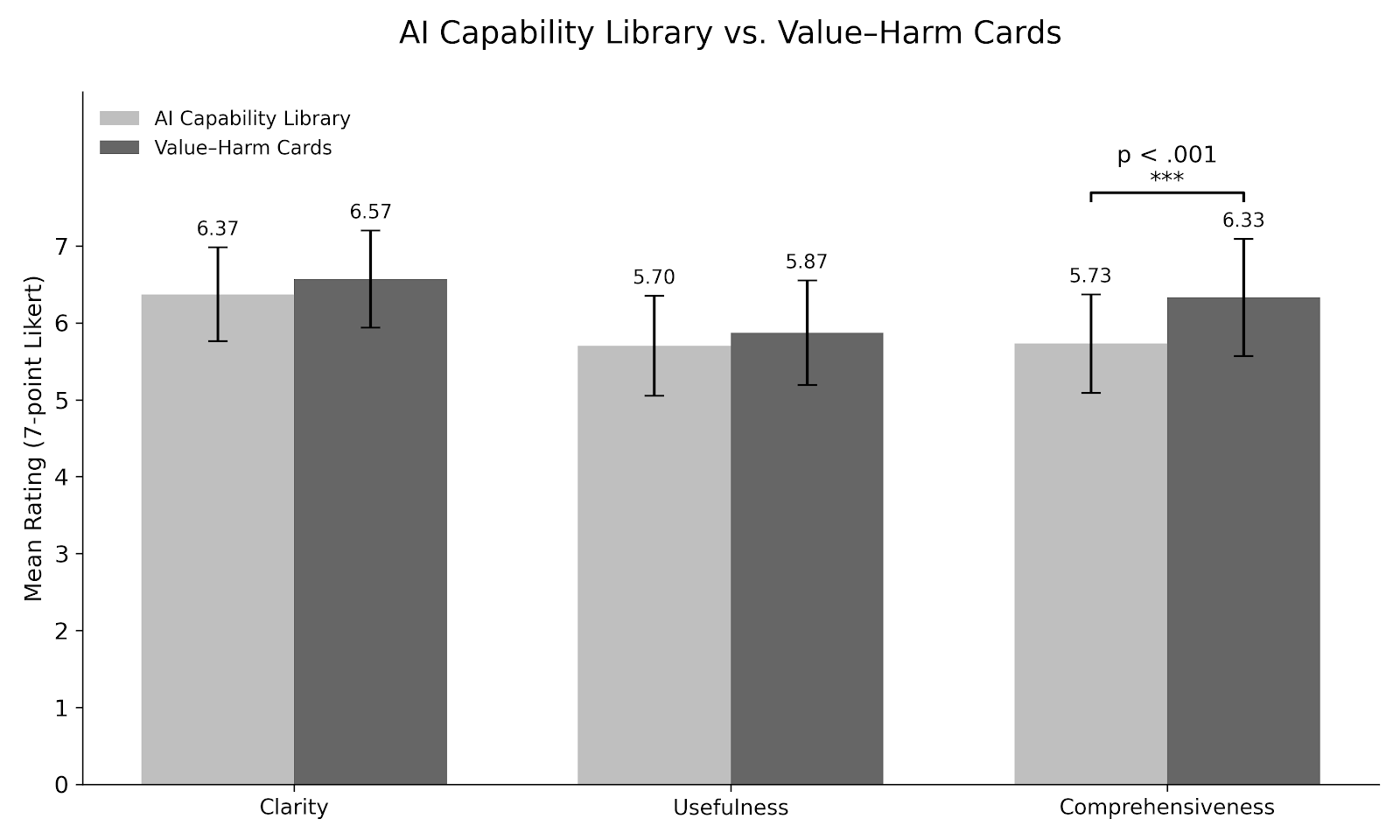}
    \caption{AI Capability Library vs. Value--Harm Cards.}
    \Description{Bar graphs for AI Capability Library vs. Value--Harm Cards, both on a 7-point Likert scale, comparing their clarity, usefulness, and comprehensiveness.}
    \label{fig:graph-library-vs-cards}
\end{figure}

Both the AI Capability Library and the Value--Harm Cards received consistently high ratings across clarity, usefulness, and comprehensiveness (\hyperref[fig:graph-library-vs-cards]{Figure~\ref*{fig:graph-library-vs-cards}}), with mean scores ranging from 5.70 to 6.57 on 7-point Likert scales. Clarity ratings were comparable for the Capability Library (\textit{M} = 6.37, \textit{SD} = 0.61) and the Cards (\textit{M} = 6.57, \textit{SD} = 0.63), \textit{t}(29) = 1.54, \textit{p} = .13, \textit{d} = 0.28. Usefulness followed a similar pattern, with the Library slightly lower (\textit{M} = 5.70, \textit{SD} = 0.65) than the Value--Harm Cards (\textit{M} = 5.87, \textit{SD} = 0.68), \textit{t}(29) = 1.07, \textit{p} = .29, \textit{d} = 0.20. Participants described both components as intuitive and straightforward, noting the clarity of capability descriptions and the actionable, example-driven design guidance.

A key divergence appeared in comprehensiveness. Value--Harm Cards were rated significantly more comprehensive (\textit{M} = 6.33, \textit{SD} = 0.76) than the Capability Library (\textit{M} = 5.73, \textit{SD} = 0.64), \textit{t}(29) = 4.09, \textit{p} < .001, \textit{d} = 0.75. Participants emphasized the cards’ capacity to spark discussion and reflection, observing that they “\textit{invite conversation}” (S08) and that the cards prompted them to “\textit{slow down}” (S04 and S16) and “\textit{reconsider what really should be designed to serve human values broadly}” (S13). The Capability Library, by contrast, was positioned as functional and referential, defining scope, articulating technical features, and supporting task-specific queries, though some noted it might need iterative updates or expanded use cases for more experienced users.

\begin{figure}[H]
    \centering  
    \includegraphics[width=\linewidth]{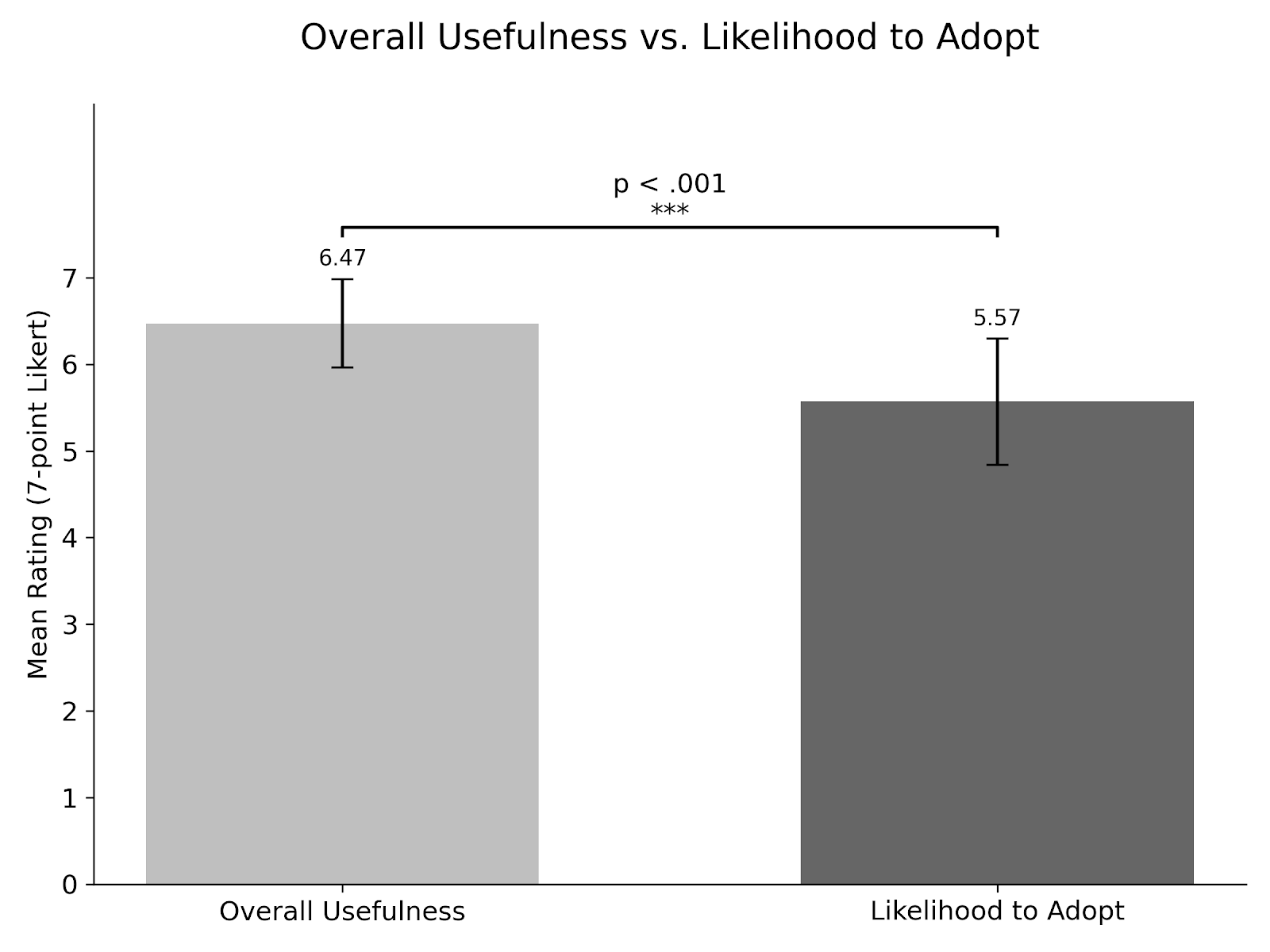}
    \caption{Toolkit’s usefulness vs. likelihood to adopt.}
    \Description{Bar graphs for Toolkit’s usefulness vs. likelihood to adopt, both on a 7-point Likert scale.}
    \label{fig:graph-usefulness-vs-adoption}
\end{figure}

This distinction became critical when evaluating overall toolkit perceptions (\hyperref[fig:graph-usefulness-vs-adoption]{Figure~\ref*{fig:graph-usefulness-vs-adoption}}). While perceived usefulness was high (\textit{M} = 6.47, \textit{SD} = 0.51), likelihood of adoption lagged (\textit{M} = 5.57, \textit{SD} = 0.73), \textit{t}(29) = 8.05, \textit{p} < .001, \textit{d} = 1.26. Correlational analysis indicated that only Value Card dimensions predicted overall impressions and adoption intentions. Toolkit usefulness correlated with Value Card usefulness (\textit{r} = .39, \textit{p} = .035) and comprehensiveness (\textit{r} = .39, \textit{p} = .034), while adoption intention correlated with Value Card usefulness (\textit{r} = .37, \textit{p} = .047), comprehensiveness (\textit{r} = .33, \textit{p} = .072), and overall toolkit usefulness (\textit{r} = .38, \textit{p} = .039). Capability Library ratings showed no significant associations (all \textit{p}s > .27), underscoring that practitioner engagement hinged on the cards’ capacity to reshape ethical reasoning.

Open-ended responses highlighted two distinct strengths that set the toolkit apart from conventional design frameworks. First, the Value--Harm Cards reframed ethical reasoning from abstract principles into negotiable, context-sensitive trade-offs. Participants emphasized that values were no longer prescriptive ideals but tensions to navigate during ideation. As S19 noted, “\textit{the cards frame values as design tradeoffs that need to be negotiated, not abstract ideals}.” Another noted the temporal shift, stating that “\textit{the toolkit seems to move ethics from a checkbox at the end to a design constraint at the beginning}” (S16). The reflective prompts introduced productive friction by surfacing ripple effects, non-user stakeholders, and long-term consequences often left implicit. Participants highlighted that contextualized prompts triggered deeper inquiry than standard design processes, prompting questions such as “\textit{which populations might be affected beyond users?}” (S13) and “\textit{what unintended consequences could emerge?}” (S11). Anchoring ethical reasoning at decision points created a space for deliberate, structured reflection before solutions became fixed.

Second, both components acted as boundary objects, facilitating shared understanding across disciplines. The Capability Library provided a common language for design, engineering, and product teams, while the Value--Harm Cards translated ethical reflection into manipulable, discussion-ready forms. S22 described the Cards as “\textit{a powerful shared language tool to align product, design, and engineering and beyond early on},” while another noted their value in legitimizing “\textit{ethical conversations with users}” (S09). Together, these components enabled documentation of not just decisions but the reasoning that underpinned them.

Despite these strengths, we identified participants’ adoption barriers among participants and provided design directions for iteration (see \hyperref[tab:barriers-directions]{Table~\ref*{tab:barriers-directions}}). A central tension emerged around the toolkit’s attempt to translate nuanced, qualitative ethical judgments into rigid, numeric scores via visual sliders, highlighting a conflict between abstraction and application. Participants (\textit{n} = 21) questioned the arbitrariness of quantification, struggling to reconcile the depth of value deliberation with the superficiality of the resulting numbers. As S10 noted, “\textit{the slider/visual evaluation definitely felt arbitrary. What’s the difference between a 5 and a 6 on Trust? I would spend more time mentally justifying the score and probably would skip it}.”

Across contexts, participants identified early-stage ideation (\textit{n} = 21) and team or stakeholder discussion (\textit{n} = 21) as prime opportunities for toolkit use. Practitioners envisioned using the toolkit in workshops to select guiding values from the outset, or as a reflective reference after initial prototypes to preempt risks. The toolkit’s structured yet flexible artifacts supported reflection that could be persisted, shared, and revisited.

\begin{table*}[t]
    \caption{Toolkit adoption barriers and design directions for future iterations.}
    \Description{Table displaying toolkit adoption barriers and design directions based on findings from phase 1 study (survey) for future iterations.}
    \label{tab:barriers-directions}
    \centering
    \small
    \begin{tabularx}{0.9\textwidth}{
    >{\raggedright\arraybackslash}p{8.25cm}
    >{\raggedright\arraybackslash}X}
    \toprule
    \textbf{Barrier} & 
    \textbf{Design Direction} \\
    \midrule
    The toolkit was designed to be a standalone web-based experience, not integrated into existing collaboration or design platforms (e.g., Miro, Figma, FigJam), creating context-switching friction in distributed work environments.
    &
    Iterate the toolkit to be embedded natively within common collaboration platforms. For example, transition the toolkit from a standalone web-based system into a design resource distributed through the Figma Community.
    \\
    \midrule
    Digital-only distribution limited tactile and collaborative interaction essential for teaching and workshop settings. The cognitive and social value of handling, arranging, and collectively manipulating cards could not be replicated by digital interfaces alone.
    &
    Provide a print-optimized PDF version, linked under the Figma Community post, to support tactile, collaborative engagement and learning.
    \\
    \midrule
    Numerical sliders intended for quantified assessment (under each value reflection card) contradicted natural collaborative practices. Practitioners negotiated ethical trade-offs through discussion rather than quantification. Sliders consumed time, introduced ambiguity, and were often bypassed in favor of qualitative dialogue, revealing misalignment with the toolkit’s reflective goals.
    &
    Remove sliders and reposition cards as primary evaluation frameworks, supporting reflection-in-action, collaborative deliberation, flexible use, and interpretive engagement rather than metric-driven assessment.
    \\
    \bottomrule
    \end{tabularx}
\end{table*}


\subsection{Design Iterations}
\label{sec:designiterations}

Guided by Phase 1 findings, we iterated on the toolkit to address key adoption barriers while preserving the qualities participants found most valuable, namely productive friction, designer agency, and support for reflective reasoning. Design changes focused on reducing workflow friction, improving pedagogical affordances, and aligning toolkit mechanisms with how practitioners naturally reason about values and trade-offs in AI concept design.

\subsubsection{From a Web-Based Tool to a FigJam Resource}

Although participants rated the toolkit as highly useful, many (\textit{n} = 26) expressed hesitation about adoption due to the need to leave their primary collaboration environments for a standalone web-based tool. To address this, we redesigned the toolkit as a FigJam resource published on the Figma Community along with a print-optimized PDF linked therein (see \hyperref[appendix:toolkit]{Appendix~\ref*{appendix:toolkit}}), so designers can access and use it directly within a platform they are already comfortable working in for ideation, collaboration, and documentation. This shift reduces context switching and supports reflection-in-action within FigJam, designers’ primary ideation environment, by allowing them to engage with the toolkit at moments when ethical deliberation naturally arise during design work, rather than in isolated sessions.

Contrary to our assumptions, participants indicated that a less interactive, more static toolkit would better support reflective thinking. Rather than automating steps or enforcing structured flows, designers valued the ability to pause, rearrange artifacts, and negotiate meaning collaboratively. In response, we intentionally reduced dynamic or system-driven interactions and designed the toolkit as a static but manipulable resource. Cards and artifacts can be freely moved, grouped, and annotated in FigJam, without enforcing a prescribed sequence of use. This design choice preserves designer agency while introducing productive friction to encourage more deliberate engagement with values and trade-offs rather than rapid completion or optimization.

\subsubsection{User Guides and Onboarding}

\begin{figure}[H]
    \centering  
    \includegraphics[width=\linewidth]{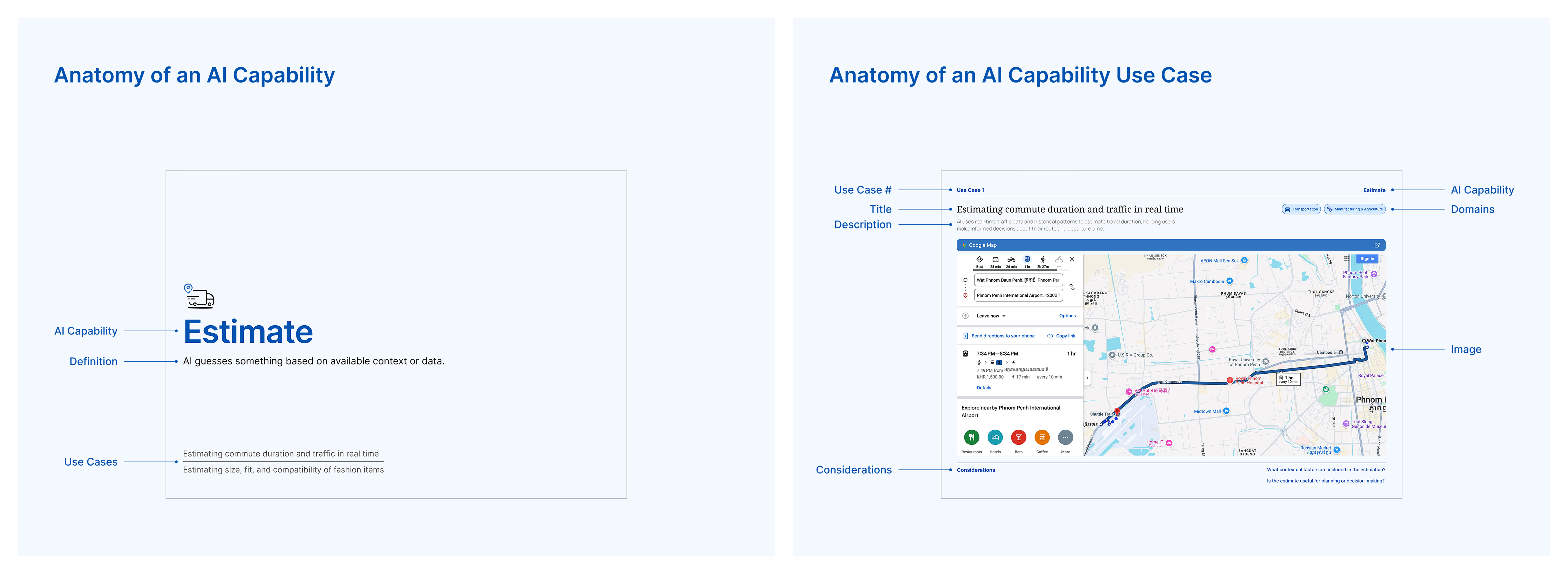}
    \caption{User guide for AI Capability Library (version 2).}
    \Description{User guide for AI Capability Library (version 2) on FigJam, showing anatomy of an AI capability, including its definition, use cases, use case number, title, description, domains, image, and considerations.}
    \label{fig:toolkit-v2-userguide-library}
\end{figure}

\begin{figure}[H]
    \centering  
    \includegraphics[width=\linewidth]{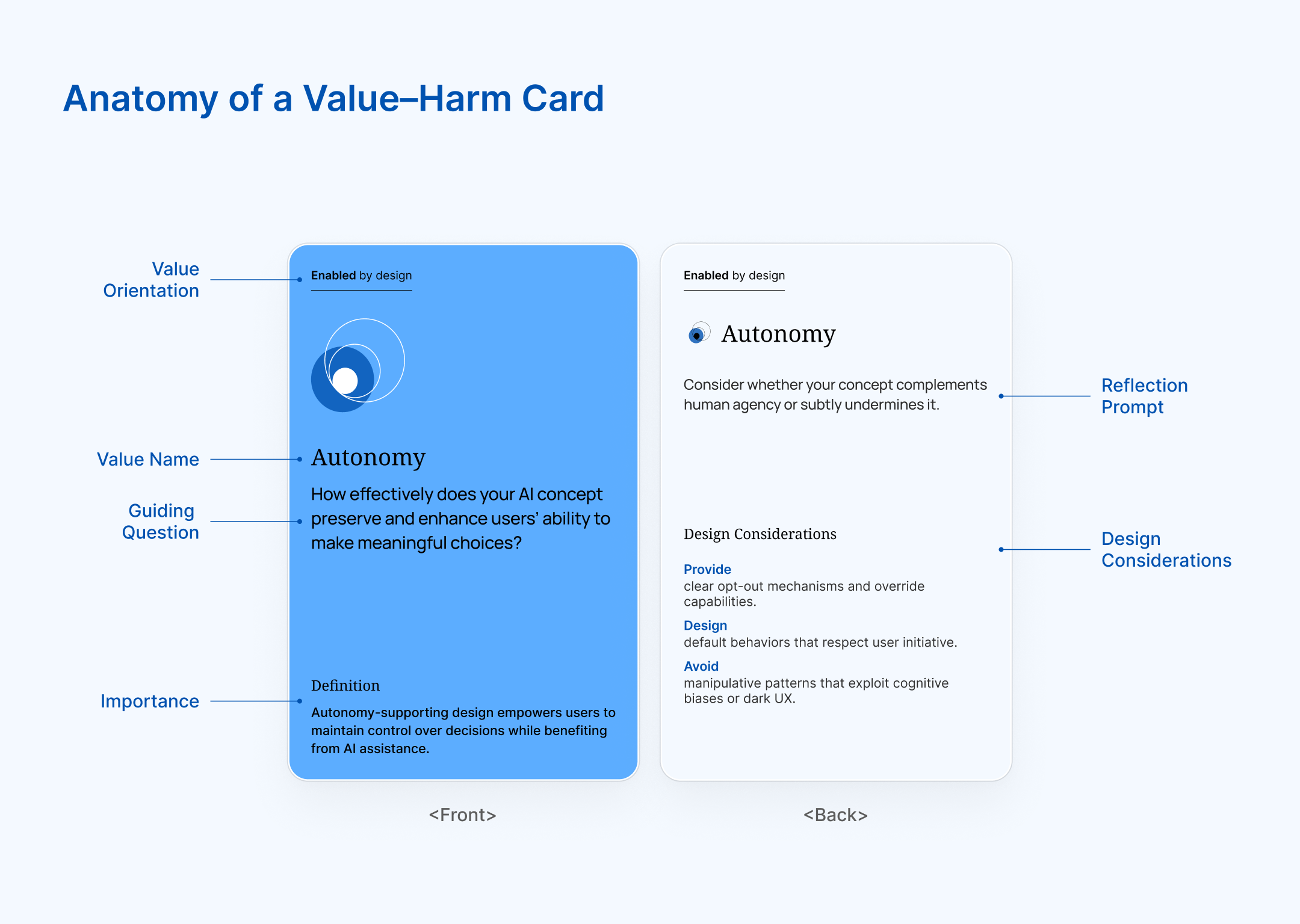}
    \caption{User guide for Value--Harm Cards (version 2).}
    \Description{User guide for Value--Harm Cards (version 2) on FigJam, showing anatomy of a Value--Harm Card, including its value orientation, value name, guiding question, definition, reflection prompt, and design considerations.}
    \label{fig:toolkit-v2-userguide-cards}
\end{figure}

To support adoption without over-structuring use, we added concise user guides that explain what the toolkit is and when it is useful (early ideation, reflection, and communication); how to browse the AI Capability Library; how to use the Value--Harm Cards; and how to use the Value--Tension Map (x/y matrix) to explore trade-offs. Moreover, the guides include an annotated anatomy of the toolkit artifacts, explaining the intent of each visual and informational element. Specifically, the anatomy of the AI Capability Library (see \hyperref[fig:toolkit-v2-userguide-library]{Figure~\ref*{fig:toolkit-v2-userguide-library}}) and Value--Harm Cards (see \hyperref[fig:toolkit-v2-userguide-cards]{Figure~\ref*{fig:toolkit-v2-userguide-cards}}) describes what each element represents and how designers might reference them during ideation and discussion. Importantly, we intentionally avoided terms such as evaluate or evaluation. Participants reported that framing design concepts as being evaluated or scored constrained creativity and discussion. Instead, the guides frame the toolkit as supporting guided reflection, positioning value consideration as an exploratory design activity rather than a judgment or assessment step.

\subsubsection{AI Capability Library Redesign}

\begin{figure}[H]
    \centering  
    \includegraphics[width=\linewidth]{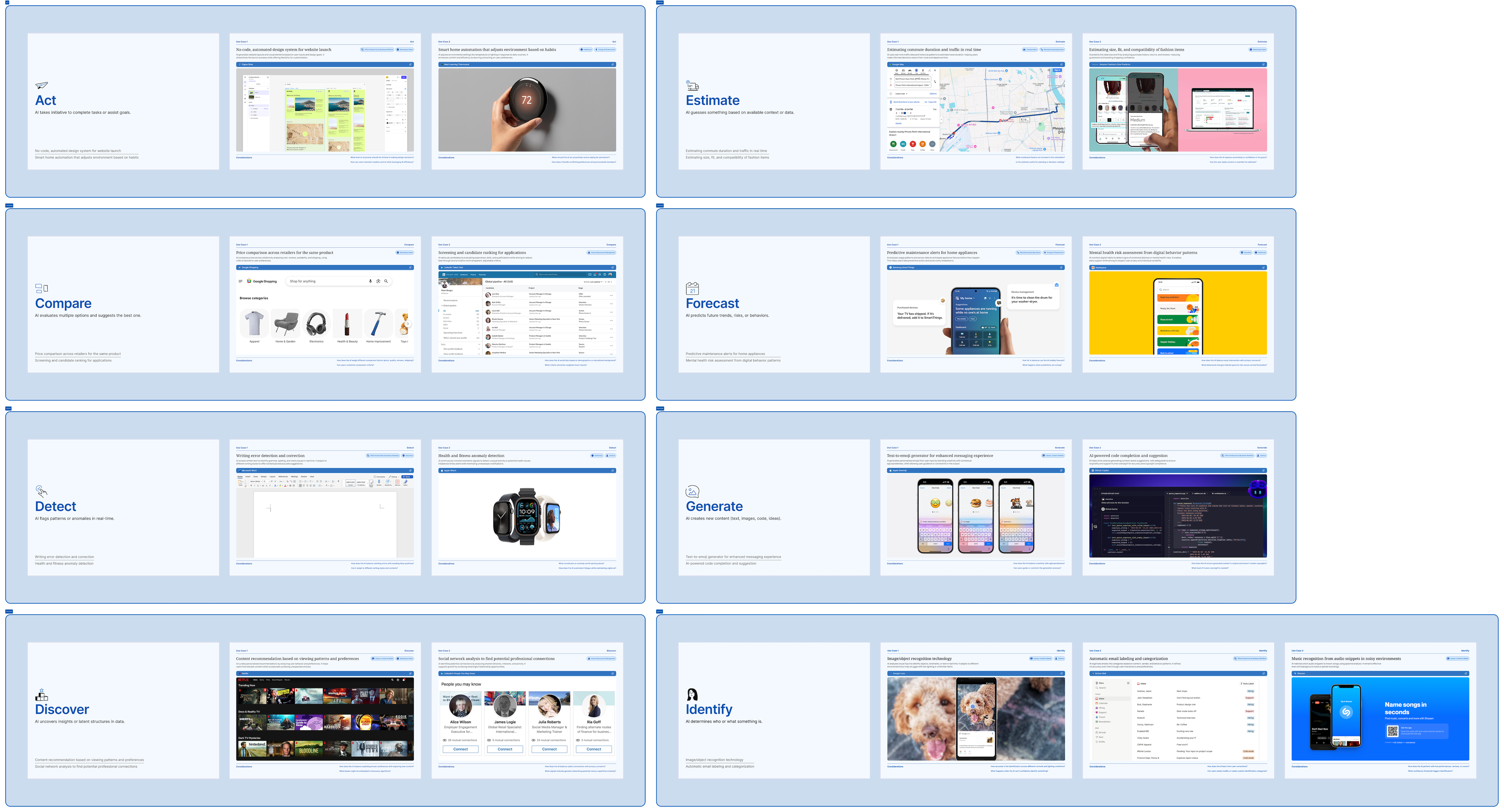}
    \caption{AI Capability Library (version 2).}
    \Description{AI Capability Library (version 2) on FigJam, showing all 8 capabilities, each with its own use cases.}
    \label{fig:toolkit-v2-library}
\end{figure}

The AI Capability Library was redesigned to maintain its role as an AI literacy scaffold when converted from a web-based interface into a static design resource. Instead of redesigning the capability definitions or adding new content, we evaluated the presentation, legibility, and functional role of the capability library in the new Figma/FigJam medium. The capabilities and their associated use cases were reorganized into a dedicated page with definitions and use case examples grouped into sections for designers to browse and visually compare (see \hyperref[fig:toolkit-v2-library]{Figure~\ref*{fig:toolkit-v2-library}}). To reduce interpretive ambiguity, each AI capability was presented using a standardized template applied consistently across the library. The template includes the capability title, description, key imagery, domains of application, design considerations, and numbered use cases.

The visualized anatomy improves legibility and coherence when the library is used as a static design resource for designers to scan, compare, and reference AI capabilities. Additionally, it reinforces the library’s role as a shared reference artifact, which supports designers in discussing feature scope and the role of AI without requiring technical knowledge. The library continues to function as an AI literacy scaffold in static form, with capability definitions and use cases presented together to support browsing and comparison.

\subsubsection{Value Reflection Undesign}

\begin{figure*}
    \centering  
    \includegraphics[width=0.85\linewidth]{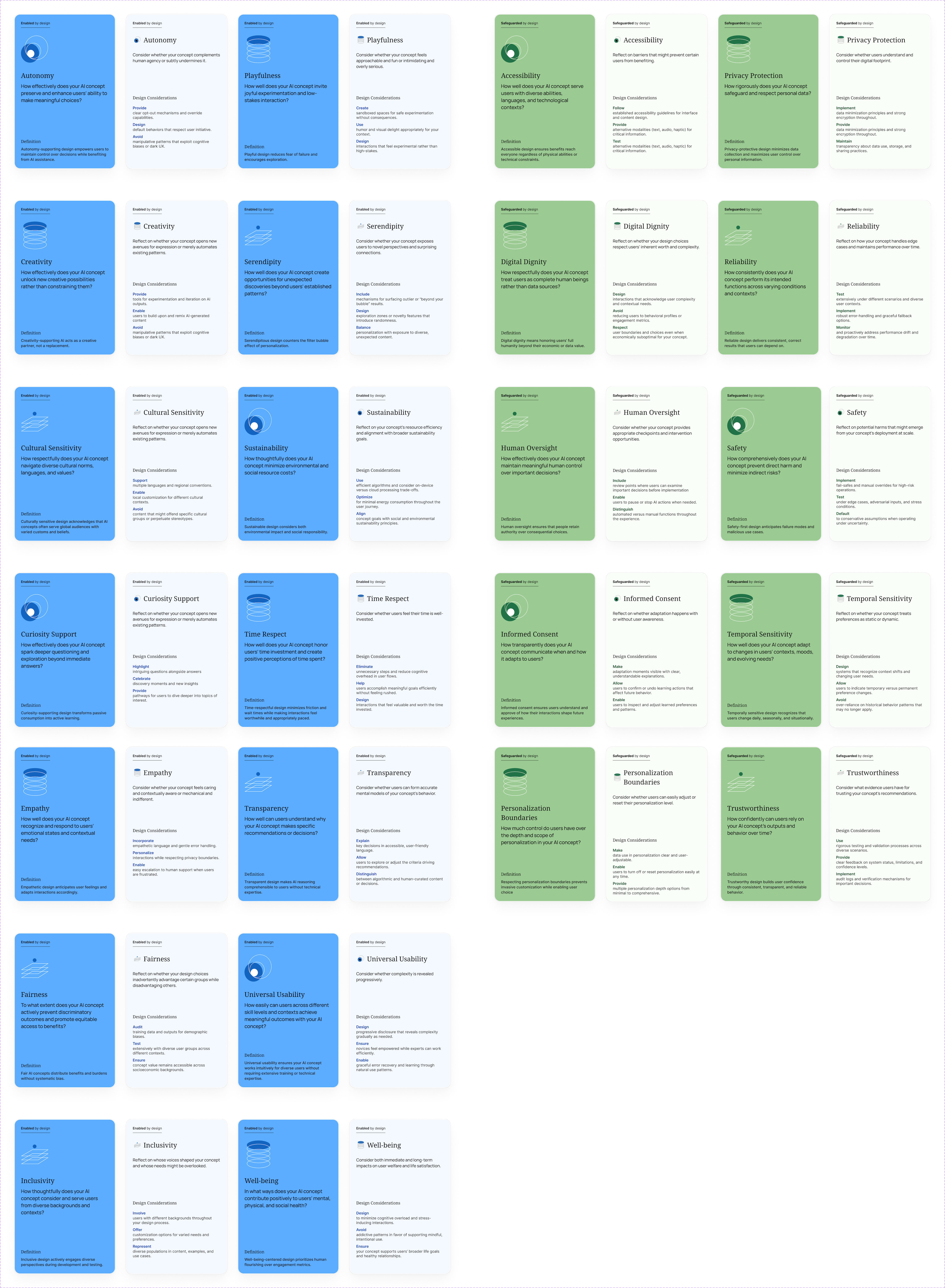}
    \caption{Full set of Value--Harm Cards (version 2) containing Enabled by Design (blue) and Safeguarded by Design cards (green).}
    \Description{Full set of Value--Harm Cards (version 2) containing 14 Enabled-by-Design (blue) and 10 Safeguarded-by-Design cards (green).}
    \label{fig:toolkit-v2-cards}
\end{figure*}

While participants valued the Value--Harm Cards for enabling collaborative discussion and negotiation, they anticipated that numerical sliders would shift attention away from ethical reasoning toward debating numbers, and that sliders would contradict how ethical trade-offs are negotiated in practice. In response, we removed the numerical sliders entirely, along with the evaluation summary report. Instead, the Value--Harm Cards were repositioned as qualitative, collaborative reflection frameworks that support reflection-in-action and deliberation rather than aggregation or comparison of scores. Because the FigJam format does not support interactive card flipping, we redesigned the card layout to clearly distinguish the \textit{front} and \textit{back} through visual cues, layout hierarchy, and labeling (see \hyperref[fig:toolkit-v2-cards]{Figure~\ref*{fig:toolkit-v2-cards}}). This ensures that prompts, descriptions, and reflective questions remain legible and interpretable in a FigJam-friendly format.

\subsubsection{Introducing the Value--Tension Map} \label{sec:valuetensionmap}

\begin{figure}[H]
    \centering  
    \includegraphics[width=\linewidth]{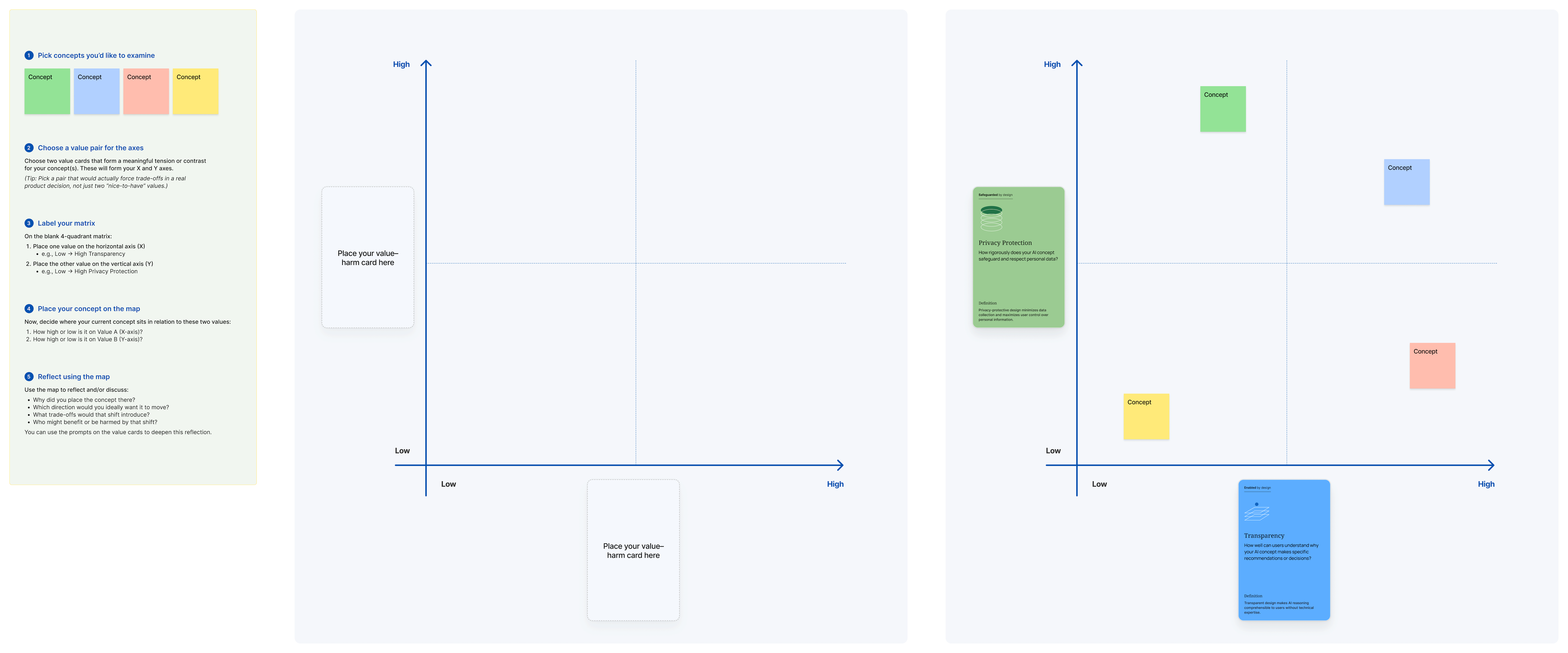}
    \caption{Instructions, Placeholder canvas, and Example of the Value--Tension Map, respectively from left to right.}
    \Description{Canvas space in FigJam for the Value--Tension Map with its instructions, placeholder canvas, and example of the Value--Tension Map with cards and concepts placed.}
    \label{fig:toolkit-v2-map}
\end{figure}

To further support collaborative reasoning, we introduced the Value--Tension Map (\hyperref[fig:toolkit-v2-map]{Figure~\ref*{fig:toolkit-v2-map}}), consisting of a blank x/y matrix and guidance for placing two Value--Harm Cards along the axes and one or more AI product concepts within the resulting quadrants. The Value--Tension Map supports relational reasoning by framing values as interdependent and negotiable rather than as isolated considerations. By positioning values along two axes, the map enables designers to reflect on trade-offs and iteratively explore alternative concept directions. 

Externalizing value tensions in a spatial form allows the map to function as a boundary object, translating abstract ethical considerations into visible, shared artifacts that support communication and collective reasoning across roles and disciplines. The map thereby enables designers to refine or select concept directions in relation to design and business priorities without relying on quantified judgment. Finally, the Value--Tension Map supports traceability by making explicit how values inform design decisions and shape downstream outcomes over time. We acknowledge that positioning values along two axes involves simplification, as tensions are not linearly measurable, and a two-dimensional layout cannot capture the full complexity of value conflicts. The map is not intended as a measurement instrument but as a means of making tensions visible and open to discussion.

\subsubsection{Visual Accessibility and Color Refinement}

As part of preparing to publish the toolkit as a Figma Community Resource, we reviewed the visual design of all toolkit components, including the AI Capability Library, Value--Harm Cards, Value Tension Map, and supporting materials. Card colors, background fills, and text contrast were checked against established accessibility guidelines to maintain legibility across the toolkit \cite{gilbert2025}. All components were evaluated for compliance with WCAG 2.1 Level AAA standards to ensure maximum accessibility. The colors were adjusted to improve contrast between the toolkit components, and we prioritized readability and consistency to support the toolkit’s use in collaborative and printed contexts. The accessibility review was conducted as part of adapting the toolkit into a static and reusable resource.


\section{Phase 2: Interview}

Following design iterations and Ledo et al. \cite{ledo2018}’s \textit{usage} as an HCI toolkit evaluation strategy, we sought to conduct a more in-depth investigation into how designers used and made sense of the toolkit while engaging in situated concept design tasks with explicit attention to designing with potential values and harms in mind.


\subsection{Method}

\subsubsection{Participants}

Phase 2 was conducted under the same IRB approval. In Phase 2, we conducted an in-depth interview study with 12 design professionals (see \hyperref[tab:participants-interview]{Table~\ref*{tab:participants-interview}}). Eligibility criteria required professional design experience and prior involvement, as a designer, in AI-adjacent or AI-focused product development. Recruitment drew from the same professional design networks as Phase 1, with attention to diversity across organizational size, industry domain, and design roles including UX/product design, service design, and design strategy. Prospective participants filled out a brief screener survey to confirm eligibility and access to Figma/FigJam.

\begin{table*}
    \caption{Participants' demographic information (P1–P12) and their assigned design brief.}
    \Description{Table displaying self-reported demographic data of 12 participants, including their  professional role, experience, organizational size, design domain, and assigned design brief.}
    \label{tab:participants-interview}
    \small
    \centering
    \begin{tabularx}{0.775\textwidth}{
        >{\raggedright\arraybackslash}p{1cm}    
        >{\raggedright\arraybackslash}p{3cm}    
        >{\raggedright\arraybackslash}p{1.5cm}  
        >{\raggedright\arraybackslash}p{1.5cm}  
        >{\raggedright\arraybackslash}p{2.35cm}    
        >{\raggedright\arraybackslash}X         
    }
    \toprule
    \textbf{ID} & 
    \textbf{Professional} 
    \textbf{Role} & 
    \textbf{Exp. (yrs)} & 
    \textbf{Org Size} & 
    \textbf{Design Domain} & 
    \textbf{Assigned Brief} \\
    \midrule
    P1 & Senior Product Designer & 8 & 100-10k & SaaS & Brief A (Education) \\
    P2 & UX Designer & 5 & 100-10k & Retail, B2B & Brief A (Education) \\
    P3 & Service Designer & 9 & <100 & Social Impact & Brief B (Urban) \\
    P4 & Product Designer & 4 & 100-10k & FinTech & Brief A (Education) \\
    P5 & Lead Design Strategist & 11 & >10k & Enterprise Tech & Brief B (Urban) \\
    P6 & UX Designer & 6 & 100-10k & Healthcare & Brief B (Urban) \\
    P7 & Interaction Designer & 8 & 100-10k & Gaming & Brief A (Education) \\
    P8 & Senior UX Designer & 10 & >10k & Enterprise Tech & Brief B (Urban) \\
    P9 & Design Lead & 7 & 100-10k & Enterprise Tech & Brief B (Urban) \\
    P10 & UX Designer & 2 & >10k & Enterprise Tech & Brief A (Education) \\
    P11 & Design Lead & 13 & 100-10k & Content, Media & Brief A (Education) \\
    P12 & Product Designer & 6 & 100-10k & Education & Brief B (Urban) \\
    \bottomrule
    \end{tabularx}
\end{table*}

\subsubsection{Procedure}

Based on the participants’ industry backgrounds, we assigned each participant to one of two situated design briefs (see \hyperref[tab:designbriefs]{Table~\ref*{tab:designbriefs}}) by pairing their primary domain with a different one (e.g., assigning a designer with an education background to brief B: AI \& Urban Living), prompting value-oriented thinking in a relatably human-centered yet unfamiliar context, rather than assessing domain-specific design expertise. Sessions lasted approximately 90 minutes and were structured in two main phases (see the Interview Protocol in \hyperref[appendix:interviewprotocol]{Appendix~\ref*{appendix:interviewprotocol}}), with flexible movement between them. The design task phase occupied roughly 45 minutes; the interview lasted 30--40 minutes; and the remaining time accommodated the natural pace of design work and discussion. Participants received their assigned brief in a private FigJam copy of the toolkit, and they had full access to all toolkit components: the AI Capability Library, Value--Harm Cards, and Value--Tension Map.

\begin{table*}
    \caption{Design briefs used to facilitate design activities and toolkit usage.}
    \Description{Table displaying 2 design briefs used to facilitate design activities and toolkit usage, including Design Brief A: AI for well-being and academic success and Design Brief B: AI for community connection and belonging.}
    \label{tab:designbriefs}
    \small
    \centering
    \begin{tabularx}{0.9575\textwidth}{
        >{\raggedright\arraybackslash}p{1.25cm}
        >{\raggedright\arraybackslash}X
        >{\raggedright\arraybackslash}X
    }
    \toprule
     & \textbf{A: AI for Well-Being and Academic Success}
     & \textbf{B: AI for Community Connection and Belonging} \\
    \midrule
    Domain
    & AI \& Education
    & AI \& Urban Living \\
    \midrule
    Audience
    & High School / College Students
    & Local Community Members \\
    \midrule
    Scenario
    & Think back to a time when school pressure, productivity tracking, or comparison with others affected your motivation, focus, or mental health. Students today often use AI-powered tools to manage their workload.
    & Consider how living in a city can sometimes feel isolating despite being surrounded by many people. AI technologies increasingly influence how people interact with their neighborhoods and social networks.
    \\
    \midrule
    Prompt
    & How might AI be designed to genuinely support students’ well-being and academic growth, rather than reinforce toxic productivity, unhealthy comparison, or performance anxiety?
    & How might AI be designed to nurture genuine human connection, inclusion, and a sense of belonging within urban communities, rather than merely optimizing convenience or efficiency?
    \\
    \midrule
    Task
    & Envision an AI-enabled solution that helps students feel seen, supported, and empowered in their academic journey. What would it look like to center values such as autonomy, empathy, and well-being in your design?
    & Envision an AI-enabled solution that helps urban residents build meaningful relationships and feel connected to their local community. How would your design embody values such as cultural sensitivity, inclusivity, and trustworthiness?
    \\
    \bottomrule
    \end{tabularx}
\end{table*}

Rather than traditional think-aloud instructions (i.e., describe what you’re thinking), we used situational framing, telling participants, “\textit{Imagine I’m a new team member joining the design team you’re leading. Walk me through your thinking as you work}.” This framing encouraged in-situ articulation of reasoning rather than surface-level descriptions. Participants worked in hybrid modalities, including reading briefs, paper brainstorming, browsing capability libraries on laptops, returning to FigJam to manipulate cards, and iterating on concepts offline using paper or tablets. During data collection sessions, we maintained a non-directive presence while offering gentle prompts to sustain articulation without coaching design choices. Immediately after the design activity, we conducted interviews covering toolkit interaction patterns, value reflection, harm reasoning, comparative advantage, and downstream impact thinking. Interviews focused on how each toolkit component influenced reasoning, moments when the toolkit changed concept considerations and direction, and perceived usefulness relative to other tools.

\subsubsection{Analysis}

All 12 transcripts were analyzed using a reflexive thematic analysis approach \cite{thematicanalysis-braunclarke}. The first author conducted an initial coding pass to generate descriptive codes, which were subsequently reviewed by a co-author who revisited segments where interpretations diverged. The first author then revisited the dataset in a second pass, refining codes through ongoing analytic engagement. Codes were progressively organized into candidate themes, which were iteratively developed and reshaped through sustained discussions between two authors. Given that the first author both conducted and analyzed the interviews, the analysis was guided by the methodological commitments of reflexive thematic analysis \cite{thematicanalysis-braunclarke}, foregrounding researcher subjectivity as an analytic lens rather than pursuing inter-rater reliability \cite{mcdonald2019}. The first author used memoing techniques \cite{birks2008memoing} to document evolving interpretive positions, which were further examined and critically extended through ongoing discussions with a co-author. Thematic saturation was considered reached when no substantively new codes or insights emerged across the final three transcripts.

\subsection{Findings}

Across all sessions, the toolkit produced observable shifts in how participants developed their AI product concepts. Rather than generating feature-level ideas, participants revised core concept directions in direct response to value--harm engagement, introduced design elements they had not initially planned, and articulated reasoning that explicitly referenced toolkit components. These shifts were visible during the design activity, highlighting the toolkit’s influence on concept envisioning practice.

Our findings are best understood as an interconnected set of cognitive and material shifts rather than discrete mechanisms. The toolkit influenced design thinking through an integrated system of elements that created friction and visibility. Across participants, this integration reshaped both how they framed problems and reasoned about solutions. Spatial, procedural, and reflective components reinforced one another, producing shifts in attention, deliberation, and conceptual framing that were observable during the process. Participants were not simply reasoning about values or harms in the abstract; they were learning through doing \cite{schon1983, sinlapanuntakul2024frustrating}. Placing cards, sketching ideas, and mapping tensions functioned as materialized inquiry, where reflection and creation were inseparable.

\subsubsection{Spatial Juxtaposition as Deliberative Friction}

Spatial juxtaposition of conflicting values and potential harms emerged as a particularly powerful mechanism for inserting intentional friction. All 12 participants began by jotting down ideas and sketching concepts based on their understanding of AI capabilities. When value cards were placed adjacent to harm cards, participants paused and reconsidered their assumptions. P7 described this process: 
\begin{quote}
    “I was sketching a feature where the AI automatically suggests breaks when it detects stress. It felt genuinely helpful. But placing \textit{autonomy} next to \textit{creativity support} made me realize I am removing the student’s choice in the name of helping them. That was uncomfortable. It forced me to ask what autonomy actually means here, like is it about the system deciding, or the student deciding? Once I asked that question, I could not turn back and just had to completely restructure the concept” (P7).
\end{quote}
Across all participants, concepts involving automation or potentially intrusive capabilities were frequently revised or removed within the activity. These interactions illustrate how making abstract conflicts visible and manipulable can produce immediate reflection and force designers to grapple with the implications of their choices in concrete, tangible ways. This spatial juxtaposition also prompted participants to balance multiple, sometimes competing values in their designs. P5 explained: 
\begin{quote}
    “I had designed everyone getting their own neighborhood feed. Personalized. Respectful of preference. But using the tension map made me ask whether, if everyone sees a different neighborhood, they are living together or separately. I realized I needed to change my main concept fundamentally. So, I kept personalization, but I added a \textit{commons} as a shared, non-algorithmic feed everyone sees. Then the \textit{personalization boundaries} card sits slightly on top of the \textit{cultural sensitivity} [card]. The tension map forced me to visualize that” (P5).
\end{quote} Many participants introduced shared or complementary design elements only after experiencing these value--harm collisions, demonstrating that spatial juxtaposition supported not merely constraint acceptance but creative reframing. Initial concepts often optimized for a single value, but post-mapping, designs evolved into AI product concepts structured to hold multiple, sometimes conflicting, values in concrete ways.

\subsubsection{Early Harm Engagement Reshapes Problem Framing}

Early engagement with potential harms further reshaped participants’ understanding of design problems and long-term impacts. Participants who encountered harm cards within the first ten minutes revised their concepts substantially more than those encountering harms later. P11 described the effect:
\begin{quote}
    “For the first 15 minutes or so, I was just locked in on feature mode, thinking about feature-level ideas like study buddy and peer feedback. But the two harm cards about \textit{temporal sensitivity} and \textit{human oversight} made me re-think about what happens to a student’s psyche after months of using these features? Then, it was natural to shift to designing for longer trajectories and impacts, not just features” (P11).
\end{quote}
Our observation confirmed that revisions addressing cumulative or systemic impacts appeared only after harm engagement. Early awareness of potential harms encouraged participants to situate design decisions within broader social and ethical contexts, and produced concepts that accounted for long-term consequences rather than immediate functionality alone. Quantitatively, early-harm participants made more substantive revisions during the activity (\textit{M} = 3.26) than late-harm participants (\textit{M} = 1.84). Substantive revisions were defined as changes to core concept direction rather than merely feature adjustments. This difference highlights that temporal positioning of harm cues can meaningfully influence the scope and depth of reflective design work.

\subsubsection{Operationalizing Values into Actionable Design Decisions}

The toolkit’s enabled-by and safeguarded-by cards helped operationalize abstract values into actionable design decisions. These scaffolds prompted participants to specify how values could be realized through concrete features or system behaviors. P6 reflected that “\textit{I put ‘trustworthiness’ as a core value because I want the system to feel honest. The safeguarded-by card asked how. And I realized trustworthiness isn’t a feeling. It requires transparent data policies, audit trails, user controls, and more. Those features aren’t glamorous in a visual way. But the card forced me to specify them}.” Our analysis further revealed that 11 of 12 participants generated specific design features only after engaging with these cards. Prior to this engagement, values remained largely aspirational \cite{umbrello2021, nishal2025}; afterward, they became concrete constraints guiding both feature ideation and interaction design. The enabled-by and safeguarded-by cards effectively bridged the gap between ethical aspirations and actionable design, supporting participants in translating abstract principles into operationalized systems.

\subsubsection{Toolkit Coherence Over Individual Components}

Finally, participants consistently emphasized that the toolkit’s effectiveness arose from the integration of its elements rather than any single component. P10 explained that “\textit{the [value] cards help list values but not design them. The capability library shows what is possible but not what matters. Harm cards show what to avoid but not what to build. So, they all create a framework for moving from thinking about ‘What’s possible?’ to ‘What’s right?’ to ‘How do I design it?’}” Participants who moved fluidly between cards, the tension map, and the capability library produced more sophisticated, multi-layered designs, whereas participants who treated elements sequentially or in isolation made fewer substantive revisions. This coherence encouraged sustained reflection on complex ethical and technical tensions, producing design concepts that were actionable, ethically informed, and socially grounded. Together, these findings demonstrate that embedding structured reflection into early-stage AI concept design can transform abstract ethical considerations into operationalized, practice-oriented decisions, providing a model for value-sensitive HCI design that integrates ethics directly into design workflows from the outset.


\section{Discussion}

Designers often struggle to translate abstract human values into concrete AI concepts while anticipating potential harms. Existing resources provide limited reflective guidance on values and societal impact during early-stage ideation. Through our study, we examined how our toolkit facilitates iterative reflection, foregrounds tensions between competing values, and encourages thoughtful consideration of ethical implications. Below, we discuss how the resource supports value--harm reflection, reveals and leverages value tensions, and introduces productive friction into the design process.


\subsection{Mapping AI Concept Envisioning Toolkit Against Existing Toolkits}

To evaluate whether the AI Concept Envisioning Toolkit is \textit{better} than existing resources, we must first clarify what \textit{better} means in this context. We define the toolkit’s advantage not through a single metric but through its capacity to help designers reason about value--harm tradeoffs and anticipate downstream, systemic impacts in early-stage AI design more effectively than existing alternatives. This requires examining both what existing toolkits miss and how our design specifically addresses those gaps.

The Envisioning Cards \cite{envisioningcards, friedman2019} represent the closest design resource to our work. Grounded in VSD’s conceptual investigation phase \cite{friedman2002, friedman2019}, which explicitly prompts practitioners to surface stakeholder impacts, value tensions, and competing priorities, they are generative, card-based, and oriented toward brainstorming and reflection about tools and technologies. Our toolkit shares these commitments but extends them in three specific ways. First, whereas the Envisioning Cards address technology broadly, ours focuses on design concepts that leverage AI. It integrates an AI Capability Library that grounds value and harm reflection in concrete technical affordances (e.g., generation, forecasting, and detection) rather than abstract technological metaphors. Second, the Value--Tension Mapping makes the relational structure of value conflicts spatially visible and manipulable as a persistent design artifact, supporting iterative, documented reasoning across design sessions in ways that physical card prompting alone does not. Third, the toolkit is embedded in FigJam, one of the primary software environments that product designers already use for brainstorming and ideation, substantially reducing the context-switching costs that limit adoption of workshop-format resources \cite{morley2020, wong2023}.

Importantly, this distinction holds even when accounting for VSD’s conceptual investigation phase \cite{friedman2002, friedman2019}, which already asks designers to reason about value trade-offs and stakeholder impacts. Our contribution is not to introduce that reasoning but to provide the scaffolding that makes it actionable within the specific constraints of AI capability ideation. We do so by grounding abstract value reasoning in concrete AI behaviors, externalizing trade-offs in a manipulable spatial form, and removing the workflow barrier that prevents ethical tools from reaching designers at the moment of ideation.

While our toolkit builds on the conceptual commitments of the Envisioning Cards, integrating the concept of VSD into a practical toolkit entails trade-offs. VSD prescribes iterative, rigorous tripartite investigations, including conceptual, technical, and empirical work, which the card-based and spatial mapping format necessarily streamlines \cite{friedman2002, friedman2019}. This approach is an intentional design decision that prioritizes accessibility and workflow integration while acknowledging that the depth of engagement the toolkit affords differs from VSD’s full framework. Our toolkit should therefore be understood as scaffolding for value--harm awareness in early-stage design practice rather than a substitute for comprehensive VSD-informed design research.

We further situate our toolkit within the broader landscape of AI design resources. Other existing resources (e.g., Google’s People + AI Guidebook \cite{googlepair}, Microsoft’s HAX Toolkit \cite{microsoftguidelines}, CMU HCII’s AI Brainstorming Kit \cite{yildirim2023designresources}) excel at different tasks. For instance, the People + AI Guidebook provides excellent patterns for AI user experience once designs are relatively stable \cite{googlepair, yildirim2023pairguidebook}. The HAX Toolkit offers rigorous methods for evaluating human-centered AI systems \cite{microsoftguidelines}. The brainstorming kit supplies a taxonomy of AI capabilities \cite{jansen2023}. Whereas many prior toolkits focus on UX best practices, model evaluation, bias auditing, or capability ideation in isolation, this work brings values, harms, and AI functionalities into the same design space and scaffolds reasoning across all three simultaneously. None of the existing resources simultaneously enable all of the following. First, \textit{in terms of timing}, in-action reasoning that brings value--harm tensions into contact with AI capabilities during ideation, a form of matchmaking between ethical and technical considerations at the moment they can still shape concept direction \cite{bly1999matchmaking}. Second, \textit{in terms of mechanism}, spatial and manipulable mechanisms for externalizing value trade-offs, turning abstract ethical reasoning into a negotiable artifact that designers can think through rather than about \cite{schon1983, giaccardi2015}. Third, \textit{in terms of integration}, seamless embedding within FigJam, a design environment practitioners already use, removing the context-switching barrier that research has consistently shown prevents ethical tools from reaching designers at the moment of ideation \cite{wong2023, subramonyam2022}. \hyperref[tab:toolkit-comparison]{Table~\ref*{tab:toolkit-comparison}} below outlines where our toolkit’s advantage lies. 

\begin{table*}
    \caption{Comparison of design-led toolkits and resources for ethical and/or AI design.}
    \Description{Table displaying comparison of our AI Concept Envisioning toolkit with 8 existing design-led toolkits and resources for ethical and/or AI design.}
    \label{tab:toolkit-comparison}
    \centering
    \small
    \begin{tabularx}{\textwidth}{
        >{\raggedright\arraybackslash}p{2.5cm} 
        >{\raggedright\arraybackslash}X   
        >{\raggedright\arraybackslash}X   
        >{\raggedright\arraybackslash}X   
        >{\raggedright\arraybackslash}X   
    }
    \toprule
    \textbf{Toolkit / Resource} & \textbf{Primary Focus \& Users} & \textbf{Approach} & \textbf{Integration \& Tensions} & \textbf{Comparison} \\
    \midrule
    AI Concept Envisioning Toolkit
    & Value–harm reflection and ethical trade-offs for designers and/in AI product teams
    & Contextual, deliberative, and explicit scaffolds for sustained reflection and AI envisioning
    & Fully embedded in Figma; usable in workshops; Promotes negotiation, discussion, and visible tension; friction is designed to provoke reasoning
    & Integrates values, harms, and technical capabilities; supports active negotiation of ethical dilemmas and actionable design decisions
    \\
    \midrule
    AIxDesign Toolkit \cite{aixdesigntoolkit}
    & Design thinking for AI for designers and innovators
    & Encourages reflection via worksheets
    & Presented as a Miro resource and print-friendly PDF; Tensions inferred from scenarios but not explicitly surfaced
    & Supports general design trade-offs, not structured for explicit value-harm negotiation
    \\
    \midrule
    CMU HCII’s AI Brainstorming Kit \cite{yildirim2023designresources}
    & AI ideation and capabilities for cross-disciplinary teams
    & Exploratory, concept-driven kit, encouraging early-stage consideration
    & Workshop-based; Friction arises from ideation constraints, not explicit value trade-offs
    & Strong for exploring capabilities but minimal scaffolding for ethical reflection
    \\
    \midrule
    Envisioning Cards (VSD) \cite{envisioningcards, friedman2019}
    & Surfacing human values in design for designers, educators, researchers
    & Contextual, exploratory, generative set of cards, supporting reflection at concept stage
    & Limited; workshop or tabletop use; Tensions implicit; encourages consideration of multiple perspectives
    & Strong for early-stage ethical consideration; lacks integration with AI capabilities or sustained iterative reflection
    \\
    \midrule
    Google PAIR Guidebook \cite{yildirim2023pairguidebook}
    & Human-centered AI UX patterns for product and design teams
    & Best practice–driven and offers reflection prompts
    & Consultative; Does not structure negotiation of value conflicts
    & Useful for aligning with UX principles, limited focus on ethical dilemmas
    \\
    \midrule
    Google What‑If Tool \cite{whatiftool}
    & Scenario-based model exploration for ML engineers
    & Allows exploration but no reflective scaffolding
    & Friction arises from model constraints, not ethical negotiation
    & Useful for testing models but does not engage with human values or value-harm trade-offs
    \\
    \midrule
    IBM AI Fairness 360 \cite{ibmaifairness360}
    & Bias detection and mitigation for data scientists, auditors
    & Focus on outcome reporting
    & Separate from design tools; Ethical reasoning abstracted and does not require discussion
    & Strong for auditing fairness metrics but not suitable for early-stage concept design or iterative value reflection
    \\
    \midrule
    Microsoft HAX Toolkit \cite{microsoftguidelines}
    & Interaction pattern design for human-AI experience teams
    & Pattern and principle-driven; Provides guidance but limited structured reflection
    & Value tensions largely implicit; friction arises from design constraints
    & Supports interaction design; ethical reasoning is peripheral
    \\
    \midrule
    Situate AI Guidebook \cite{situateAIguidebook}
    & Deliberative early AI design for public sector teams
    & Question-based deliberation; Provides structured prompts for discussion
    & Friction emerges from multi-stakeholder negotiation, context-specific
    & Scaffolding is discussion-dependent; Focused on public sector
    \\
    \bottomrule
    \end{tabularx}
\end{table*}

Our research demonstrates each of these advantages through multiple forms of evidence. The design differences are substantive and documented in detail in \hyperref[sec:designiterations]{Section~\ref*{sec:designiterations}}. Harm considerations are integrated as first-class scaffolds rather than secondary checklists, values and harms are manipulable rather than static, and the toolkit operates within rather than alongside designers’ daily environments. Empirical evidence from Phase 2 shows that these design choices produce observable changes in designer reasoning. Designers encountered value conflicts at moments that forced concept rethinking. Harm cards encountered early reshaped problem definitions, not just added constraints \cite{jung2025, namer2025harm96designers}. Designers specified values more concretely through scaffolding that pushed them from aspiration to concept implementation. Most importantly, these shifts happened in real time, during ideation, not after designs had solidified \cite{schon1983, schon1987, nelson2014}.

Phase 2 findings make this temporal difference concrete. P11, for instance, spent the first fifteen minutes in feature mode before harm cards about temporal sensitivity and human oversight redirected their thinking toward longer trajectories. Without those cards present at that moment, the feature-level framing would have persisted. These questions became unavoidable within the activity rather than surfacing later through critique or testing, when feature commitments would already have been made \cite{yang2020}. Designers encounter these questions during ideation, when they have maximum flexibility to reshape their concepts. This temporal difference is foundational. Existing toolkits largely accept that ethical reasoning happens late; our toolkit argues it should happen early, and provides the scaffolding to make that feasible \cite{liao2023, friedman2019}.


\subsection{Scaffolding Value--Harm Reflection}

Existing resources for ethical AI design typically present principles, heuristics, or checklists that make values explicit but offer limited support for situated reflection during early, open-ended concept work \cite{googlepair, ibmguidelines, morley2020}. These approaches often assume that clarity and specificity are sufficient for ethical engagement but leave designers on their own when grappling with uncertainty about how values relate to concrete design decisions. This limitation is particularly salient in early ideation, where ambiguity is not a failure of knowledge but a generative condition for creative reasoning \cite{buchanan1992, cross2021}. By focusing scaffolding on both values and harms, designers are supported in navigating this ambiguity as part of concept envisioning rather than treating it as a separate compliance exercise \cite{abedin2022}.

Turning values and harms into manipulable artifacts is the mechanism that makes this possible. Prior approaches summarized in Table~\ref{tab:toolkit-comparison} tend to frame ethics as prescriptive, implying a single “correct” interpretation that does not reflect the diverse ways values manifest in different contexts \cite{whittlestone2019, umbrello2021}. In contrast, the toolkit asks designers to articulate trade‑offs between values, to align them with concrete AI capabilities, and to trace how both values and harms might play out across stakeholders and use situations. Because the reflection is situated in the context of uncertain design possibilities, designers are pushed to think through implications rather than simply check against a predetermined list of criteria \cite{liao2020, benjamin2021}, which turns ambiguity from a barrier into a resource for deeper reasoning \cite{tian2024, dorst2015}.

Evidence from our study suggests that this scaffolding also strengthens collaboration. Ethical reasoning is often treated as an individual cognitive effort, resulting in fragmented understanding across disciplines \cite{deng2023, nahar2022, kross2021}. By making trade-offs and potential harms visible in shared artifacts that teams can jointly manipulate, the toolkit supports negotiation of meaning, alignment of priorities, and collective documentation of rationale \cite{mao2019, subramonyam2022}. This distributed sensemaking addresses a persistent gap, as current practice lacks tools that support both explicit value articulation and collective interpretation during uncertain early design stages \cite{muralikumar2025, xu2019, shneiderman2020hcai}.

Finally, linking design decisions explicitly to articulated trade-offs and harm considerations supports sustained ethical reflection beyond any single session. The record of reasoning generated through this process creates continuity across iterations, enabling teams to revisit and refine both their concepts and their value judgments over time \cite{schon1992, raji2020}. This sustained engagement surpasses checklists and aligns ethical reflection with how designers revisit, revise, and reframe ideas throughout the lifecycle of a concept \cite{buxton2010, dorst2015, cross2021}.


\subsection{Revealing and Leveraging Value Tensions}

Ambiguity in concept work is not just inevitable; it is central to how designers explore novel possibilities \cite{buchanan1992, cross2021, nelson2014}. Yet many structured approaches to ethical AI design simplify or sidestep this ambiguity through linear workflows or numeric evaluations that constrain generative reasoning. Making value tensions explicit reframes these tensions not as obstacles but as resources that can inform both ethical reasoning and creative exploration \cite{sadek2024designing, shen2024valuecompass}. Artifacts that externalize conflicting priorities allow teams to explore alternatives, compare interpretations, and examine systemic consequences in ways that summary scores or rigid heuristics do not \cite{doordan2003, giaccardi2015, dove2020}. Our study surfaced several such patterns. Designers initiated ideation by foregrounding values, paired AI capabilities with ethical considerations, and used prompts to trace downstream effects, illustrating how surfacing tensions can shape idea generation in practice \cite{palani2024, rayan2024}.

Leaving room for interpretation is key to this mechanism. Conventional approaches equate usability with clarity, often reducing reflection to simplified metrics or fixed evaluation steps that give the illusion of certainty. In contrast, allowing ambiguity in the representation of value tensions introduces a form of productive challenge as designers have to articulate assumptions, weigh trade-offs directly, and engage with conflicting priorities rather than gloss over them \cite{benjamin2021, tian2024}. In early conceptual spaces, such deliberate uncertainty allows teams to test multiple potential pathways instead of converging prematurely on a single solution \cite{kumar2024}. In this way, ethical and creative reasoning can proceed in tandem \cite{anderson2024, fu2024, crilly2019}. Translating this insight into practice, however, requires deliberate facilitation. Design teams accustomed to converging quickly on solutions may find unresolved value tensions disorienting without structured guidance on how to work through conflict productively rather than around it. The Value--Tension Mapping addresses this in part by giving ambiguity a visible, manipulable form around which teams can reason.

When value tensions are made visible and remain open to interpretation, collaboration across disciplines becomes more tractable. Ethical reasoning is often tacit and poorly shared across technical, design, and managerial roles, leading to misunderstandings and misaligned expectations \cite{kross2021, zhang2020}. Structuring reflection around ambiguous but visible tensions creates a shared space for negotiation, enabling teams to co-construct interpretations, document rationale, and align decisions without stifling creative exploration \cite{shi2023, muralikumar2025}.

The emergent practices observed show that surfacing value tensions under ambiguity can transform early conceptual exploration into an ethically informed, generative process. Rather than reacting to prescriptive guidelines after ideas are already formed, designers in our study used contradictions between values and capabilities to reveal new directions and possibilities. This pattern demonstrates that engaging openly with value conflicts can broaden the creative scope of concept ideation \cite{crilly2019, verganti2020, benjamin2021}.


\subsection{Designing for Productive Friction}

Productive friction \cite{cox2016friction, wakkary2016friction} is the toolkit’s organizing principle. Many existing ethics tools minimize cognitive load, inadvertently suppressing the reflective engagement complex problems require \cite{baron2015, abedin2022, yang2020}; our toolkit instead introduces friction through open-ended prompts and artifacts that resist premature closure. Doing so encourages designers to slow down, reflect, and interrogate trade-offs \cite{wakkary2016friction}, particularly during early conceptual exploration, which is inherently undetermined \cite{schon1983, buxton2010}.

This friction works because it preserves agency while providing scaffolding. Prescriptive workflows or automated evaluations often dictate a single sequence of steps, encouraging designers to locate the \textit{right answer} quickly and move on. In contrast, a friction-based approach leaves designers free to choose when and how they engage with value tensions \cite{popa2021, nelson2014}. By sustaining ambiguity and resisting enforced convergence, the toolkit supports iterative exploration of alternatives and keeps ethical considerations active throughout development rather than relegating them to validation after the fact \cite{friedman2019, liao2023, shneiderman2020bridging}.

Introducing friction also encourages systemic thinking \cite{cox2016friction}. Our prompts and Value--Tension Map push designers to consider indirect and long-term consequences, extending ethical reflection beyond immediate user concerns \cite{verbeek2011, brundage2020, rahwan2019}. Designers often assume such reasoning will happen naturally, but our observations show that without structural support, teams consistently overlook broader impacts \cite{namer2025harm96designers, saxena2025}.

Finally, productive friction supports traceability and iterative improvement. Interrupting workflows at key reflective moments produces shared artifacts and documented rationale that persist beyond single sessions, enabling concepts to evolve while preserving their ethical grounding \cite{schon1992, raji2020}. This approach embeds reflection into the core of concept envisioning rather than treating it as an optional add-on. By embracing ambiguity and the productive challenges it poses, this framing addresses persistent gaps in prior work and advances approaches for responsible, reflective AI design \cite{shneiderman2020hcai, shneiderman2020bridging}.

Our findings are grounded in work with design professionals, yet AI products are developed by cross-functional teams whose members each make consequential decisions. The productive friction that designers found generative may function differently in roles where professional norms prioritize speed and technical precision over careful reflection. Furthermore, because we structured our sessions to afford time specifically for using the toolkit, productive friction is not unconditionally beneficial in typical commercial design sprints. In fast-paced design environments, intentional interruptions to the design momentum may be experienced as burdensome, particularly when teams face tight organizational timelines.


\subsection{Reflection on RtD Process}

We reflect on our process not merely as a sequence of design decisions but as an experimental space in which we learned how tools, values, and ethical reasoning interact in practice \cite{zimmerman2014RtD, bardzell2015}. Engaging collaboratively with our four team members, we noticed that what initially appeared as ambiguity was not a deficit but a generative space. Early in the process, we assumed designers would prefer clear, prescriptive guidance for \textit{good} AI design. Through iterative reflection, we recognized that ambiguity created productive tension, compelling practitioners to negotiate meaning within their contexts, reconcile conflicting priorities, and surface assumptions that would otherwise remain invisible. We reflect on this as a core insight that ethical reflection in design cannot be fully scripted but emerges through structured engagement with uncertainty \cite{schon1983, schon1992, cross2021}.

We reflect on the tensions we navigated between familiarity and friction. Initially, we worried that a toolkit unfamiliar to designers might hinder adoption, but we observed that the intentional friction of confronting value conflicts and societal impacts was essential for meaningful reflection. Designers’ discomfort was not a flaw but an affordance, prompting deeper reasoning and conscious decision-making. Similarly, we reflect on temporal tensions. While the AI Capability Library provided immediate utility, orienting designers quickly, the Value--Harm Cards supported sustained engagement, prompting reflection that persisted across sessions and projects. This indicates that designing for value-oriented AI requires scaffolds that support both immediate understanding and engagement over time.

Our reflections on values were equally revealing. We initially treated values as constraints or ethical checkboxes, but through our process we observed that practitioners appropriated them as creative levers. Selecting values before ideation reframed the design process, allowing values to guide concept generation rather than being retrofitted post hoc. Values also proved contextual and relational; ethical reasoning shifted depending on stakeholders and systemic implications, revealing that values are never universal absolutes but situated guides that must be interpreted and negotiated. We also observed the personal dimension of values. Designers engaged emotionally, connecting professional judgment with individual ethical perspectives, which sustained commitment and reflection beyond abstract principles \cite{nishal2025, nakata2025}.

We reflect on the patterns that emerged from these engagements as indicative of alternative paradigms for ethical design practice. Value-first ideation, ripple reflection rituals, and capability-value pairing were not imposed but surfaced organically from practitioners’ interactions with the toolkit. These patterns show that ethical reflection is performative; it happens through doing, negotiating, and iterating, rather than through abstract instruction alone. We also reflect on the toolkit as embedded educational infrastructure, supporting ongoing learning and accommodating diverse entry points depending on expertise or project needs. Through these experiences, values became ongoing infrastructure rather than one-off considerations, shaping not just individual sessions but the entire design practices and future projects.

Finally, we reflect on what our process taught us about RtD itself. The iterative, co-creative, and reflective approach allowed us to uncover both expected and unexpected insights, highlighting the entanglement of tools, ethical reasoning, and practice. We recognize that supporting value-oriented AI innovation is less about providing answers or prescriptive workflows and more about creating conditions for sustained engagement, where designers are prompted to reflect, negotiate, and act in context. Our reflections underscore that the design of ethical AI is inseparable from the design of reflective practice. Design tools must scaffold engagement, preserve productive tension, and enable designers to treat values as generative materials rather than constraints \cite{sinlapanuntakul2026envision}.


\section{Limitations and Future Research} 
\label{sec:limitations}

While this research demonstrates the toolkit’s capacity to support early-stage value reasoning, several limitations warrant acknowledgment. First, the toolkit was developed and evaluated primarily within Western, technology-industry design contexts. Cultural, organizational, and disciplinary differences can shape how values, harms, and ethical reasoning are understood and prioritized. Systematic validation across more diverse contexts is needed to assess whether these mechanisms generalize.

Second, our Phase 1 evaluation relied on single-item measures for each dimension, limiting psychometric rigor and construct validity compared to validated multi-item scales. The uniformly high ratings likely reflect the composition of the sample, specifically experienced designers with AI project experience, and may not generalize to the broader range of practitioners who would encounter the toolkit in practice. Future evaluations should develop validated multi-item measures to enable more robust comparison of how tools support reflective depth, value operationalization, and anticipatory harm reasoning.

Third, all participants in both phases were design professionals presumably oriented toward reflective practice. Cross-functional teams (e.g., product managers, data scientists, and engineers) who build AI-enabled products and systems may require different scaffolding or vocabulary, and the productive friction designers found generative may function differently under professional norms that prioritize speed and precision over careful reflection. Evaluation with interdisciplinary teams is therefore an important direction for future research. Moreover, the toolkit does not address the role of datasets in AI concept design, given that training data encodes assumptions about representation, inclusion, and harm that are inseparable from the value implications of AI capabilities. Future iterations could explore how dataset provenance and curation choices might be surfaced within the AI Capability Library, where capability definitions implicitly depend on the quality and composition of underlying data.

A natural future direction is to extend the toolkit with a complementary set of design principles. While the toolkit currently integrates value reflection, harm consideration, and AI capability grounding into a single proactive framework, its use remains episodic and context-dependent. Design principles would allow designers and product teams to reason about potential harms and ethical trade-offs even when the toolkit itself is not actively in use, reinforcing structured and actionable reasoning throughout early-stage AI concept design at scale.


\section{Conclusion}

Concept envisioning is a critical point of leverage in AI design, yet designers often lack integrated resources to engage with values and potential harms at this stage. This paper presents the AI Concept Envisioning Toolkit as a contribution toward addressing this gap. Developed through an RtD approach, the toolkit combines an AI Capability Library, Value--Harm Cards, and a Value--Tension Map to support reasoning about technical feasibility, values, and harms during ideation. Our findings show that embedding value reflection directly into existing design workflows helps designers surface value tensions earlier, anticipate unintended harms, and make more substantive revisions before concepts solidify. Rather than treating ethics as a downstream concern, the toolkit introduces productive friction by externalizing values and harms into shared, discussion-ready artifacts that support collective sensemaking and reasoning beyond what abstract principles alone can achieve. For HCI researchers and practitioners, our findings suggest that embedding value reflection at moments of concept formation can advance ethical AI design by making value trade-offs explicit, actionable, and embedded in everyday design practice.


\begin{acks}
We thank Axel Roesler and Gary Hsieh for their feedback on the toolkit design and our Phase 2 plan, and Rupal Patel for her involvement in early concept exploration and ideation.

\paragraph{Toolkit Contributions} Pitch Sinlapanuntakul directed the conceptual vision and component-level decisions, under which Soyun Moon, Yuri Kawada, and Yeha Chung contributed equally to the initial toolkit conceptualization. Soyun led the design iterations and user guide development for the Value--Tension Map and Value--Harm Cards; Yuri refined the visuals and user guide for the AI Capability Library; and Yeha refined the visual design of the Value--Harm Cards and ensured accessibility compliance.
\end{acks}


\bibliographystyle{ACM-Reference-Format}
\bibliography{references}


\onecolumn
\appendix


\section{Toolkit}
\label{appendix:toolkit} 
Our toolkit is publicly accessible via the following links:
\begin{itemize}
    \item Toolkit (FigJam resource) on Figma Community: https://bit.ly/ACEtoolkit
    \item Toolkit (PDF version): https://bit.ly/ACEtoolkit-pdf
\end{itemize}


\section{Survey Protocol}
\label{appendix:surveyprotocol}

\subsection{Part 1: AI Capability Library}
Participants watched a short walkthrough video of the AI Capability Library and rated its perceived value on 7-point Likert-scale items with some open-ended questions.

\begin{enumerate}
    \item Clarity of capability presentation \\ 
    (1 = Not at all clear; 7 = Very clear)
    \item Please explain your above rating
    \item Usefulness of example use cases \\ 
    (1 = Not at all useful; 7 = Very useful)
    \item Please explain your above rating
    \item Comprehensiveness of the capability range \\ 
    (1 = Very inadequate; 7 = Very comprehensive)
    \item Please explain your above rating
    \item Explain how the library could support your design practice
    \item Suggestions for AI Capability Library improvement
\end{enumerate}

\subsection{Part 2: Value--Harm Cards}
Participants watched a short walkthrough video of the Value--Harm Cards and rated the component’s perceived value across the same three, 7-point Likert-scale, followed by some open-ended questions.

\begin{enumerate}
    \item Clarity of card design including layout and examples \\ 
    (1 = Not at all clear; 7 = Very clear)
    \item Please explain your above rating
    \item Usefulness of reflective prompts for critical reasoning \\ 
    (1 = Not at all useful; 7 = Very useful)
    \item Please explain your above rating
    \item Comprehensiveness of human and societal considerations represented (1 = Very inadequate; 7 = Very comprehensive)
    \item Please explain your above rating
\end{enumerate}

\subsection{Part 3: Overall Toolkit Assessment}
\begin{enumerate}
    \item Overall usefulness (1 = Not at all useful; 7 = Very useful)
    \item Likelihood of adoption (1 = Very unlikely; 7 = Very likely)
    \item Most valuable toolkit components (select all that apply)
    \begin{itemize}
        \item AI Capability Library
        \item Value Reflection Cards
        \item Sliders / Visual Evaluation
        \item Other (open-ended)
    \end{itemize}
    \item Please explain your above choice(s)
    \item Design stages that you envision using this toolkit (select all that apply)
    \begin{itemize}
        \item Early ideation and exploration
        \item Mid-stage evaluation
        \item Final design assessment
        \item Team/stakeholder discussion
        \item Other (open-ended)
    \end{itemize}
    \item Please explain your above choice(s)
    \item Advantages this toolkit offer over your current practice (open-ended)
    \item Challenges/barriers to adoption (open-ended)
    \item Suggestions for toolkit improvement (open-ended)
\end{enumerate}


\section{Semi-Structured Interview Protocol}
\label{appendix:interviewprotocol} 

\subsection{Concept Summary}
\begin{enumerate}
    \item Walk me through the final AI concept you just envisioned. 
    \item What problem does it address and what values does it attempt 
    to serve?
    \item As you designed this/these concept(s), did you find yourself thinking about downstream effects---how your design might shape users’ behaviors weeks or months from now? 
\end{enumerate}

\subsection{Toolkit Usage}
\begin{enumerate}
    \item Walk me through the order in which you engaged with the toolkit elements. Which did you engage with or reference first, and why?
    \begin{itemize}
        \item[] \textit{Probes: When did you turn to value--harm cards? Did the AI 
        capability visuals influence what features you imagined?}
    \end{itemize}
    \item Walk me through a moment when you interacted with one toolkit component that directly changed your concept direction (if any).
    \begin{itemize}
        \item[] \textit{Probes: Was it a specific enabled-by/safeguarded-by card, the tension map, or a specific AI capability? Walk me through where that change happened.}
    \end{itemize}
    \item How did the toolkit influence your thinking/ideation processes during the activity?
\end{enumerate}

\subsection{Value--Harm Reflection}
\begin{enumerate}
    \item How did the toolkit help you reason about values for your concept(s)? Did any card surface a value you wouldn’t normally have articulated? How so?
    \item How did you identify harms, and how did the toolkit support that? Did seeing harms alongside values change your reasoning compared to considering them separately?
    \item The toolkit separates enabled-by design from safeguarded-by design cards. How did that distinction influence your thinking?
    \item Walk me through how you used the value--tension mapping. How did pairing values or visualizing trade-offs influence your thinking?
\end{enumerate}

\subsection{Comparative Advantage}
\begin{enumerate}
    \item Compared to other design toolkits or frameworks you have used, what does this toolkit offer that feels genuinely different? How so?
    \item If you were to use only one toolkit element---just the 
    capability library, or just value--harm cards---would it be as useful as having all the pieces together? Why or why not?
\end{enumerate}

\subsection{Suggestions}
\begin{enumerate}
    \item What one change you think might increase the toolkit’s effectiveness for concept envisioning with values and harm?
\end{enumerate}


\end{document}